\documentclass[twocolumn]{aastex62}
\usepackage{lineno,hyperref,amsmath}

\shorttitle{RATs and alignment of irregular grain ensemble}
\shortauthors{Herranen, Lazarian, and Hoang}

\watermark{DRAFT}
\setwatermarkfontsize{200pt}

\begin{document}

\title{Radiative torques of irregular grains: Describing the alignment of a grain ensemble}

\correspondingauthor{Joonas Herranen}
\email{joonas.herranen@iki.fi}

\author[0000-0001-7732-9363]{Joonas Herranen}
\affiliation{Department of Physics, University of Helsinki}

\author[0000-0002-7336-6674]{A. Lazarian}
\affiliation{Department of Astronomy, University of Wisconsin}

\author[0000-0003-2017-0982]{Thiem Hoang}
\affiliation{Korea Astronomy and Space Science Institute}
\affiliation{Korea University of Science and Technology}

\begin{abstract}
The radiative torque (RAT) mechanism is the most promising way of explaining observed polarization arising from aligned grains. We explore the efficiency of the grain alignment by an anisotropic radiation flow for an extensive ensemble of grain shapes. We calculate the distribution of the ratios of the amplitudes of the two major components of the RATs, that is an essential parameter that enters the theory of RAT alignment in Lazarian \& Hoang (2007, LH07). While this distribution is different for different classes of grain shapes that we considered, the most probable values of the parameter are centered in the range of $q^{max}\sim 0.5-1.5$. The functional form from RATs calculated is in good agreement with the analytical model (AMO). We find that the RAT efficiency scales as $(\lambda/a)^{-3}$ for $\lambda\gg a$ as previously found in LH07. This increases the power of predictions obtained with the RAT theory. We also confirm that superparamagnetic inclusions are necessary in achieving high degrees of alignment, and constrain the parameter space describing the requirements for achieving these alignment degrees.
\end{abstract}

\section{Introduction} \label{sec:intro}

Polarization from aligned grains is both an important informant about magnetic fields in diffuse media and molecular clouds as well as is a major impediment in the search for the enigmatic cosmological B-modes. The mystery of grain alignment has been one of the astrophysical problem of the longest standing with the first detection of aligned grains reported in \citet{Hiltner1949} and \citet{Hall1949}. And the first theories of grain alignment suggested shortly after that \citep{Davis1951, Gold1952a}. Later quite a number of processes of grain alignment have been discussed in the literature (see \citet{Lazarian2003b} for a review) with the legends of astronomy, e.g. Ed Purcell and Lyman Spitzer, working intensively on the problem. Their work clarified many key astrophysical processes, but could not provide the theory of alignment that could explain the polarization observations. Apparently, the approaches explored missed a key element. This missing element was introduced in the work by \citet{Dolginov1976a}, which was published in relatively obscure journal of Soviet Astronomy. This prophetic work suggested that some irregular grains are helical in terms of their interaction with the radiation, i.e. that they can scatter different amounts of right and left radiation. As a result, such grains subject to anisotropic radiation are expected to get spun-up and aligned. The requirements for the grains to be helical were not defined by the authors, neither the analytical calculations for the given shape that they adopted as an example of a helical grain were confirmed by the subsequent studies.\footnote{In fact, the adopted for their calculations shape was was too much symmetric to produce any torques \citep{Hoang2009}.} Nevertheless, \citet{Dolginov1976} introduced the new idea that changed the direction of the subsequent grain alignment research. Another important lesson from the \citet{Dolginov1976} study was that the approximation of grains by spheroids was missing the essential pieces of physics. 

The next step in the grain alignment saga was done by \citet[henceforth DW96]{Draine1996a} where the efficiency of radiative torques was calculated using the advanced Discrete Dipole Scattering (\textsc{ddscat}) code by \citet{DDSCAT}. This work for the first time demonstrated  the strength of radiative torques. In particular, it was shown that for typical ISM conditions the radiative torques can spin up grains up to the rotational velocities in excess of any other spin up mechanism, e.g. mechanism of spin up related to the formation of molecular hydrogen over the dust grain surface. The latter was suggested in \citet{Purcell1979} and was considered the dominant process of dust suprathermal rotation. While this work of DW96 brought the radiative torques in the spotlight of astrophysical research, its deficiency was that for the calculations of the estimates of the grain rotation were obtained ignoring the dynamics of the grains in the beam of radiation. Thus DW96 provided the upper limits of grain rotation that the radiative torques can induce. For the alignment mechanism DW96 assumed the classical \citet[see also \citet{Spitzer1979,Purcell1979}]{Davis1951} paramagnetic relaxation mechanism. This were only consistent for the isotropic radiative torques which, however, were $\sim 100$ times weaker than the anisotropic torques of the radiation beam.

The problem of the radiative torque alignment induced by anisotropic radiation was addressed in the subsequent study by \citet[henceforth DW97]{Draine1997}. That study pioneered many elements that were used in the research that followed. In particular, to describe the complex dynamics of grains subjected to a beam of radiation the phase trajectories of grains were traced and the attractor and repellor points were calculated. This makes possible to observe the outcome of the complex grain dynamics as grains interact with anisotropic radiation. The deficiency of this study, however, was that the crucial element of grain dynamics, i.e. the crossover, was disrgarded in the analysis. The crossover takes place as the grain slows down so the value of the angular momentum perpendicular to the grain axis of the maximal moment of inertia gets comparable with the value of angular momentum parallel to the axis of the maximal moment of inertia. The theory of crossovers was suggested by \citet{Spitzer1979} and was extended in \citet{Lazarian1999a}\footnote{\citet{Lazarian1999a} took into account that the thermal fluctuations within the grain material \citep{Lazarian1994a,Lazarian1997b} and this changed the crossover dynamics, in particular the degree of stochastic randomization from gaseous bombardment during the crossover}. Without the treatment of crossovers in DW97, the obtained grain dynamics was distorted, e.g the cyclic trajectories were reported for most of the numerical tests. These trajectories happened to be an artifact of the adopted model. The subsequent study in \citet{Weingartner2003} addressed the issue of crossovers and reported the existence of the attractor points corresponding to a very low angular momentum. This study did not resolved the issue with the cyclic trajectories which also presented in the study.

The above studies of the radiative torques provided the foundations for the further progress of the theory. First of all, it became clear that radiative torques were an important element of the grain alignment theory and they should not be disregarded. The ways of calculating the value of radiative torques and the dynamics of grains were introduced. The deficiency of these studies was that the reason for grain alignment remained unclear and the quantitative theory of grain alignment was still missing. The alignment of grains that corresponded to observations, i.e. with long axes perpendicular to magnetic field, was tested for a few shapes and the reason why this alignment happens preferentially to the opposite type of alignment, i.e. the alignment with long grain axes parallel to magnetic field, was unclear. 

The set of problems above was addressed in \citet[henceforth LH07]{Lazarian2007b} where the analytical theory of grain alignment was suggested. The study in LH07 returned to the original idea in \citet{Dolginov1976a} that to experience radiative torques the grains should have intrinsic helicity. However, it proposed a model radically different from the latter study. The Analytical MOdel (AMO) in LH07 employed a macroscopic toy model of a helical grain, which was an oblate grain with an mirror attached to it at 45 degrees to it (see Figure 1, "toy model in LH07", to be added). The calculations of the torques provided in the framework of geometric optics, nevertheless provided a remarkable correspondence with the shape of the radiative torques obtained with the \textsc{ddscat} for most of the wavelength corresponding to the interstellar spectrum. For instance, the model explained the significant differences in the functional dependences of the torques for some of the numerically explored grain shapes. The radical change of the torque shape originated from the helical grains being either left-handed or right-handed. With the AMO, the grain alignment became a predictive theory. LH07 used the abbreviation for RAdiative Torques, i.e. RATs, which became an accepted term \citep[see][]{Andersson2015} which we will use for the rest of the paper. 

To compare the RATs arising from the AMO and from the \textsc{ddscat} calculations LH07 provided the comparison of torque components in the lab frame, which was different from the earlier studies. In this frame the component $Q_{e3}$ (see Figure \ref{fig:coords}) was found to be present even for spheroidal grains for which both $Q_{e1}$ and $Q_{e2}$ components vanish. The component $Q_{e3}$ was identified with the cause of grain precession in along the direction of radiation anisotropy.\footnote{When this precession is faster than the Larmor precession, the grain gets aligned in respect to radiation rather than the ambient magnetic field. This can be termed "k-RAT" alignment as opposed to the "B-RAT" alignment in respect to magnetic field. The B-type alignment is typical for the interstellar medium, while in the vicinity of bright sources, e.g. stars, novae, supernovae the alignment can happen in respect to the radiation direction (LH07). This provides an interesting way of measuring magnetic field strength or/and grain magnetic response \citep{Lazarian2018}.}  The role of components $Q_{e1}$ and $Q_{e2}$ was explored in LH07 and they were identified as the cause of the RAT alignment. Their functional dependence, i.e. dependence on the angle between the radiation anisotropy direction and the grain axis of maximal moment of inertia (see Figure 1) was shown to be similar to that of AMO. However, the ratio of the amplitude value of the torque components, i.e. $q^{\mathrm{max}}\equiv Q_{e1}^{\mathrm{max}}/Q_{e2}^{\mathrm{max}}$ was changing from one irregular grain shape to another. LH07 study showed that the properties of alignment, i.e. the alignment with low angular momentum or high angular momentum for a given direction between the direction of magnetic field and the radiation anisotropy, depends on the parameter $q^{\mathrm{max}}$. This made $q^{\mathrm{max}}$ an essential parameter of the RAT theory. While LH07 provided the calculations of $q^{\mathrm{max}}$ for the 3 shapes in DW96 plus additional 2 shapes, the total number of the explored shapes amounted only to 5, which precluded any quantitative conclusions of what the expected distribution of $q^{\mathrm{max}}$ one should expect. 

The current study addresses the deficiency of the theory above and explores the distribution of $q^{\mathrm{max}}$ parameters for a collection of grain shapes of different classes. Our goal is (1) to provide an insight of what to expect of the collection of arbitrary shaped grains and (2) provide a way to limit the distribution of grain shapes on the basis of polarization observations. 

In what follows, in Sections \ref{sec:RATdesc} and \ref{sec:dustmodel}, we briefly describe the numerical scattering solution and the dust model which are used as the basis of this work. In Section \ref{sec:analysis}, we study the RAT properties of different ensembles and compare the results to those of AMO. We discuss our results in Section \ref{sec:discussion}, and present our conclusions in Section \ref{sec:conclusions}.

\section{RATs for irregular grains}\label{sec:RATdesc}
In order to obtain intuition about the alignment of irregular grains via any quantity describing alignment, two practical issues must be addressed. First, as the properties of interstellar dust are not scrictly known, a large set of candidate analogues are to be considered. Second, as RAT alignment is inherently a repeated, dynamically changing scattering problem, an efficient numerical solution is essential. In this section, these issues are considered.

\subsection{The $ T $-matrix solution of radiative torques}
The $ T $-matrix method, originally formulated by \citet{Waterman1965}, describes electromagnetic scattering in a concise manner. The shape, composition and size information of the scatterer is encoded in the $ T $-matrix, which maps the incident radiation into scattered radiation the the vector spherical wave function expansion. 

As the properties of the scatterer are described strictly by the $ T $-matrix, with no dependence on the incident radiation direction of beam shape, a repeated solution of scattering by an unchanging scatterer is efficient in the $ T $-matrix formulation. In recent years, methods of finding the $ T $-matrix of arbitrary scatterers have been developed. 
This makes it a viable method when considering alignment of irregular grains. In this work, the $ T $-matrices are determined via a volume integral equation approach \citep{Markkanen2012,Markkanen2016}. 

Generally, under anisotropic radiation fields, the radiative torque $ \mathbf{\Gamma}_{\mathrm{rad}} $ can be defined as 
\begin{equation}
\mathbf{\Gamma}_{\mathrm{rad}} = \dfrac{\bar{u}_{\mathrm{rad}}a_{\mathrm{eff}}^{2}\bar{\lambda}}{2}\gamma\mathbf{Q}_{\Gamma}(\Theta,\beta,\Phi),
\end{equation}
where the radiation environment is described by its radiation anisotropy  degree $ \gamma $, the mean wavelength $ \bar{\lambda} $ and mean energy density $ \bar{u}_{\mathrm{rad}} $, $ a_{\mathrm{eff}} $ is the equivalent volume sphere radius of the grain, and $ \mathbf{Q}_{\Gamma} $ is the torque efficiency. In the $ T $-matrix framework, the radiative torque $ \mathbf{\Gamma}_{\mathrm{rad}} $ can be directly written in an analytical form, in a sense as a function of the total fields \citep{Farsund1996}.

\subsection{Describing the RAT alignment}
\begin{figure}
	\centering
	\includegraphics[width=\linewidth]{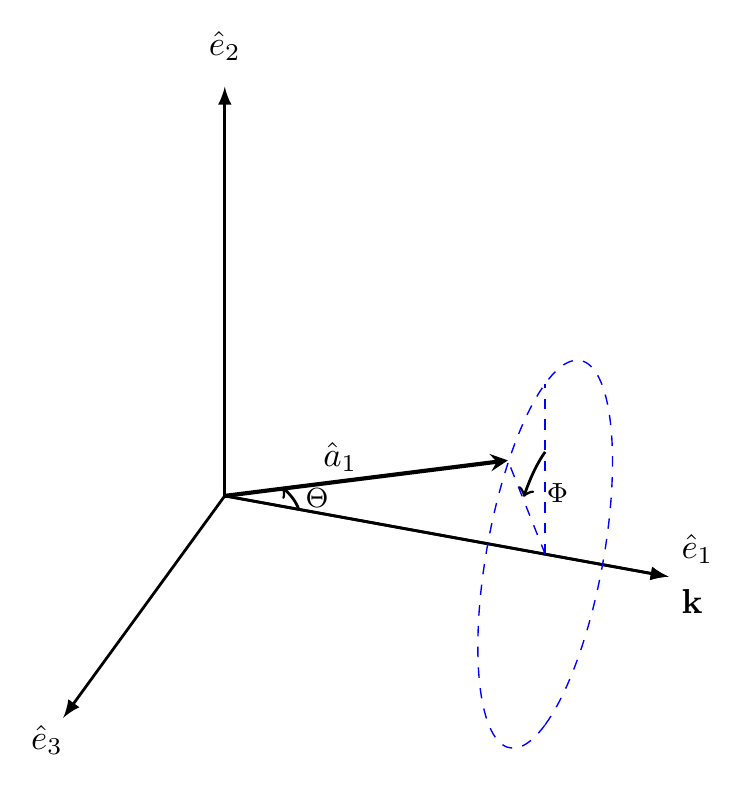}
	\caption{Scattering (laboratory) coordinate system, in which $ \beta $-averaging is done around $ \hat{a}_{1}(\Theta,\Phi) $. }
	\label{fig:coords}
\end{figure}

An important quantitity describing grain alignment is the ratio $ q^{\mathrm{max}} $, or the $ q $-factor. The $ q $-factor is given by $ Q^{\mathrm{max}}_{ei} $, $ i = 1,2 $, which are maximal magnitudes of the $ \beta $-averaged RAT efficiency components in the scattering coordinates. The scattering coordinate system allows a decomposition of RAT efficiency as
\begin{equation}
\begin{aligned}
\mathbf{Q}_{\Gamma}&(\Theta,\beta,\Phi) =  Q_{e1}(\Theta,\beta,\Phi)\mathbf{\hat{e}}_{1} \\
&+ Q_{e1}(\Theta,\beta,\Phi)(\mathbf{\hat{e}}_{2}\cos\Phi + \mathbf{\hat{e}}_{3}\sin\Phi) \\
& +Q_{e1}(\Theta,\beta,\Phi)(\mathbf{\hat{e}}_{3}\cos\Phi - \mathbf{\hat{e}}_{2}\sin\Phi).
\end{aligned}
\end{equation}
The coordinate system is illustrated in Figure \ref{fig:coords}. For the remainder of this work, RATs are calculated averaging over $ \beta $ and setting $ \Phi = 0$. The $ q $-factor has been identified as an important measure of alignment in LH07. The dominantly aligning component affects what kind of attractor points the grain has in its alignment phase space. Namely, when $ Q_{e1} $ dominates, or $ q^{\mathrm{max}} $ is larger than unity, grains have to high-$ J $ attractor points at $ \Theta = 0 $. Otherwise the points are repellors, and the only attractors are low-$ J $ at $ \Theta = \pi/2$. In addition, when magnetic fields are included, the low-$ J $-only alignment is very probable for grains with $ q^{\mathrm{max}} \approx 1 $. 

\begin{figure}
	\centering
	\includegraphics[width=\linewidth]{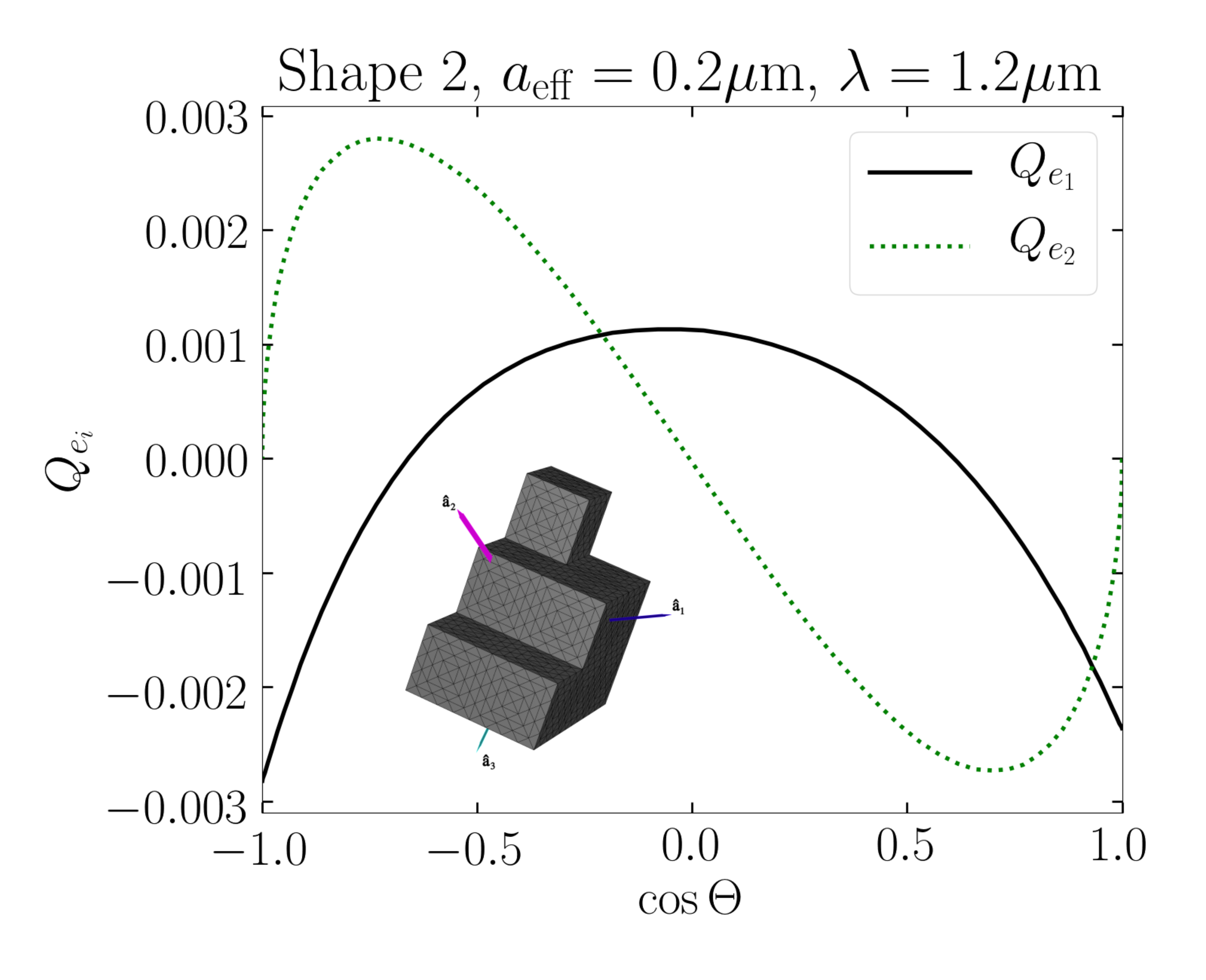}
	\caption{Reproduction by the $ T $-matrix method of the rotation-averaged radiative torques from \cite{Lazarian2007b}.}
	\label{fig:Qshape2}
\end{figure}

The $ T $-matrix method can reproduce previous \textsc{ddscat} results in LH07, which is illustrated in Figure \ref{fig:Qshape2} for a shape known as "Shape 2", composed of "astronomical silicate" \citep{Draine1984}. The maximal values of radiative torque components give $ q^{\mathrm{max}} = 1.007$. 

\section{Model dust grain ensemble}\label{sec:dustmodel}
\begin{figure}
	\centering
	\includegraphics[width=\linewidth]{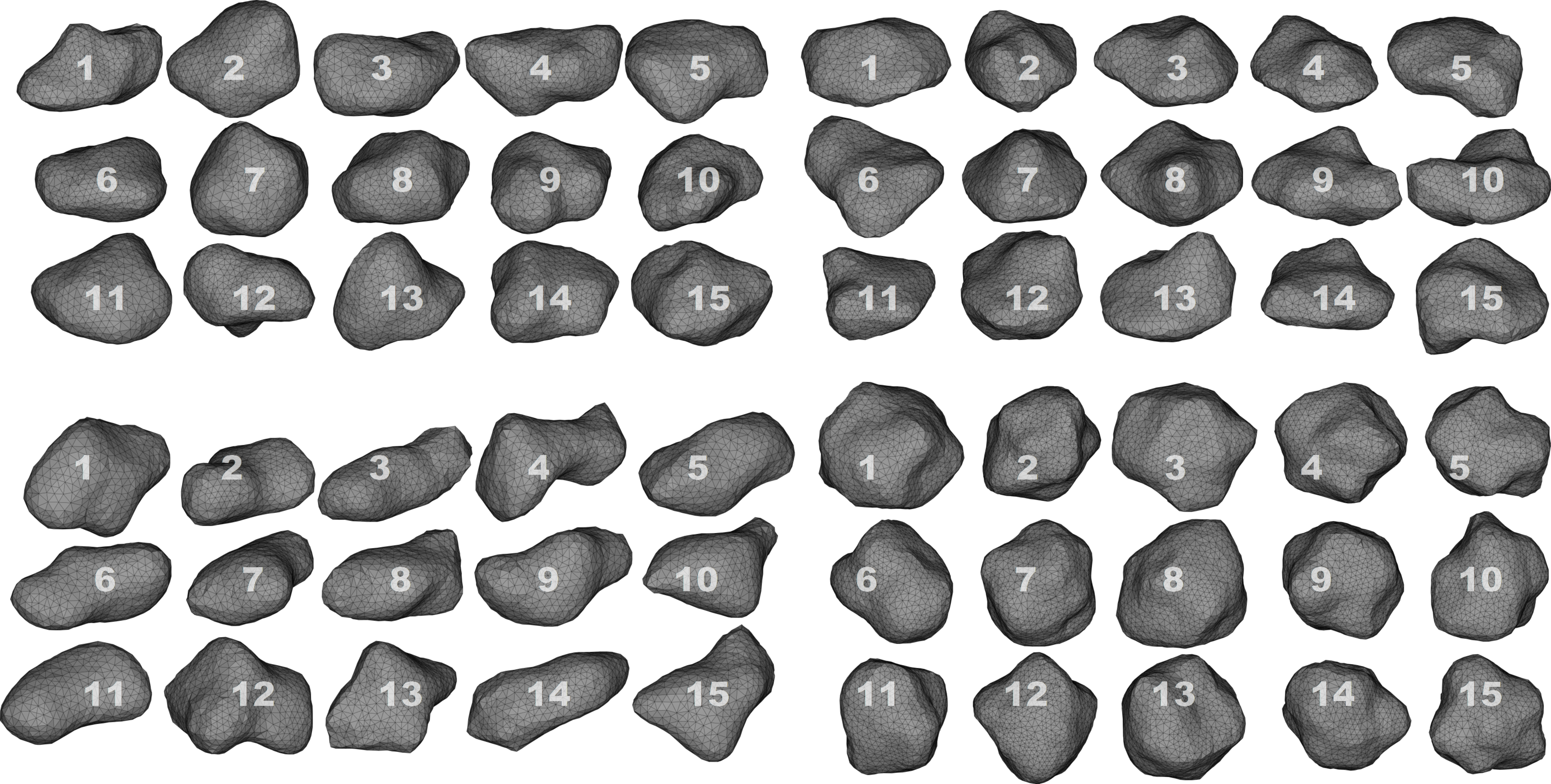}
	\caption{Gaussian ellipsoids, oblate spheroids, prolate spheroids, and spheres, respectively from upper left to bottom right in numbered groups of 15. Numbering also corresponds to generating indices of the shapes and will be used to identify particles later in the text.}
	\label{fig:shapes}
\end{figure}

The $ T $-matrices have been calculated for 4 ensembles of Gaussian random ellipsoids \cite{Muinonen2011}, with their base shapes being spherical, ellipsoidal, oblate spheroidal and prolate spheroidal. The generating parameters are summarized in Table \ref{table:shapetable}. Each base shape is deformed 15 different times, resulting in 60 total particle shapes. The shapes are illustrated in Figure \ref{fig:shapes}.

\begin{table}
	\centering
	\caption{The generating parameters, axial ratio, standard deviation $ \sigma $, and correlation length $ l $, for each base shape. 15 different deformed shapes are generated with a seed index 1--15 for each base shape.}
	\begin{tabular}{r|ccc}
		Base shape & Axial ratio & $ \sigma $ & $ l $\\
		\hline 
		Ellipsoid & 1:0.8:0.6 & 0.35 & 0.125 \\ 
		Prolate spheroid & 1:0.5:0.5 & 0.35 & 0.125 \\ 
		Oblate spheroid & 1:1:0.5 & 0.35 & 0.125 \\ 
		Sphere & 1:1:0.99 & 0.35 & 0.125
	\end{tabular} 
	\label{table:shapetable}
\end{table}

For each shape, the $ T $-matrices are calculated for 10 different wavelengths (linearly spaced between 0.120 and 1.740 $ \mu $m), 3 different particle sizes ($ a_{\mathrm{eff}} $ = \{0.05, 0.1, 0.2\} $ \mu $m) and 10 different compositions. In this work, a composition of pyroxine-type silicate with a small mantle of carbonaceous material is considered to make up all the grains, unless otherwise stated. The optical properties are available from \citet{Jones2017} and references therein.

\begin{figure}
	\centering
	\includegraphics[width=\linewidth]{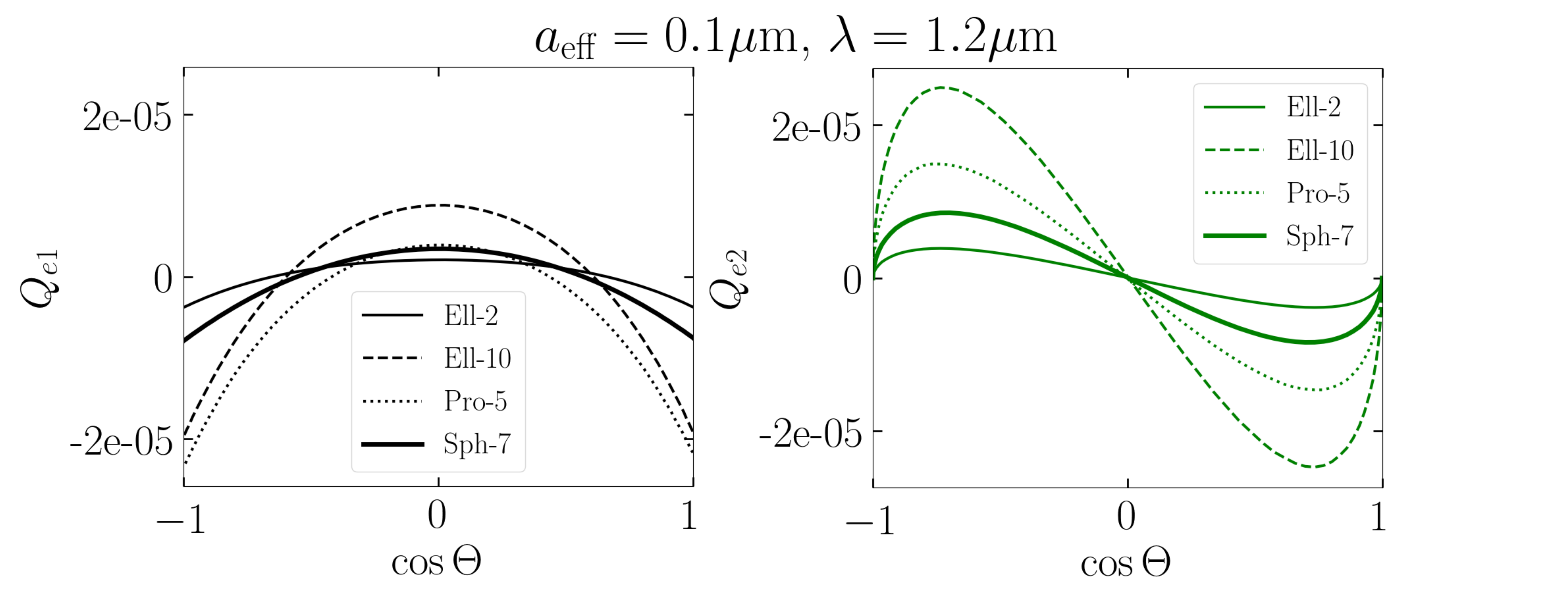}
	\caption{The left-handed RAT components $ Q_{e1} $ and $ Q_{e2} $ for 5 random coated silicate shapes for $ \lambda = 1.200 $ $ \mu $m.}
	\label{fig:1200nmRAT-lhand}
\end{figure}

\begin{figure}
	\centering
	\includegraphics[width=\linewidth]{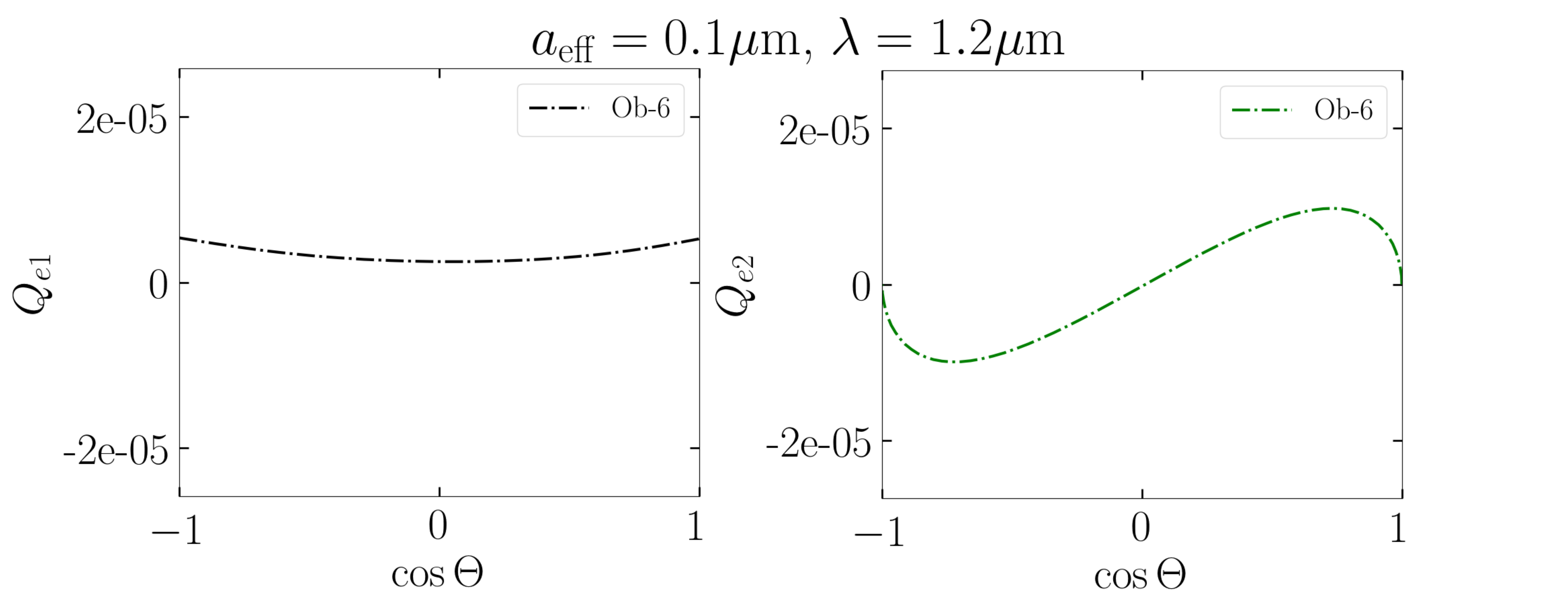}
	\caption{The same as in Fig. \ref{fig:1200nmRAT-lhand}, but for the right-handed components.}
	\label{fig:1200nmRAT-rhand}
\end{figure}

\begin{figure}
	\centering
	\includegraphics[width=\linewidth]{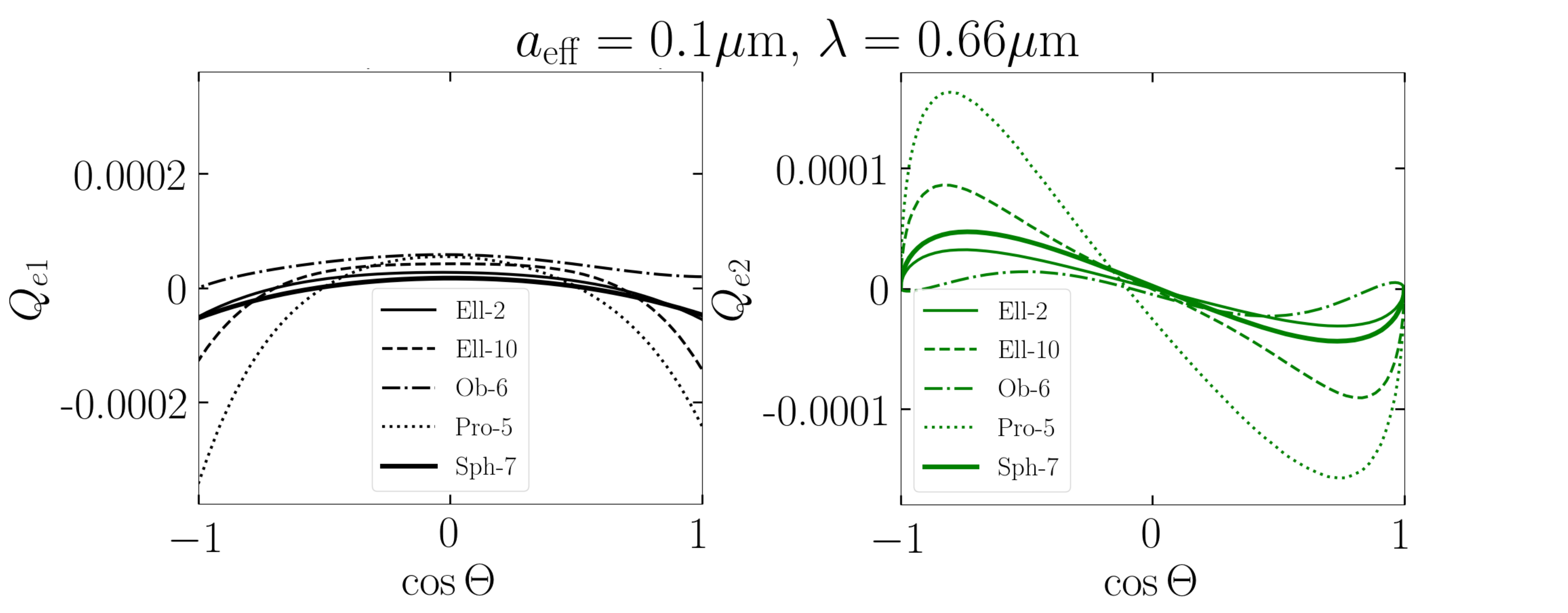}
	\caption{The same as in Fig. \ref{fig:300nmRAT-lhand}, but for $ \lambda = 0.660 $ $ \mu $m. All grains show left-handed helicity in this case.}
	\label{fig:660nmRAT-lhand}
\end{figure}

\begin{figure}
	\centering
	\includegraphics[width=\linewidth]{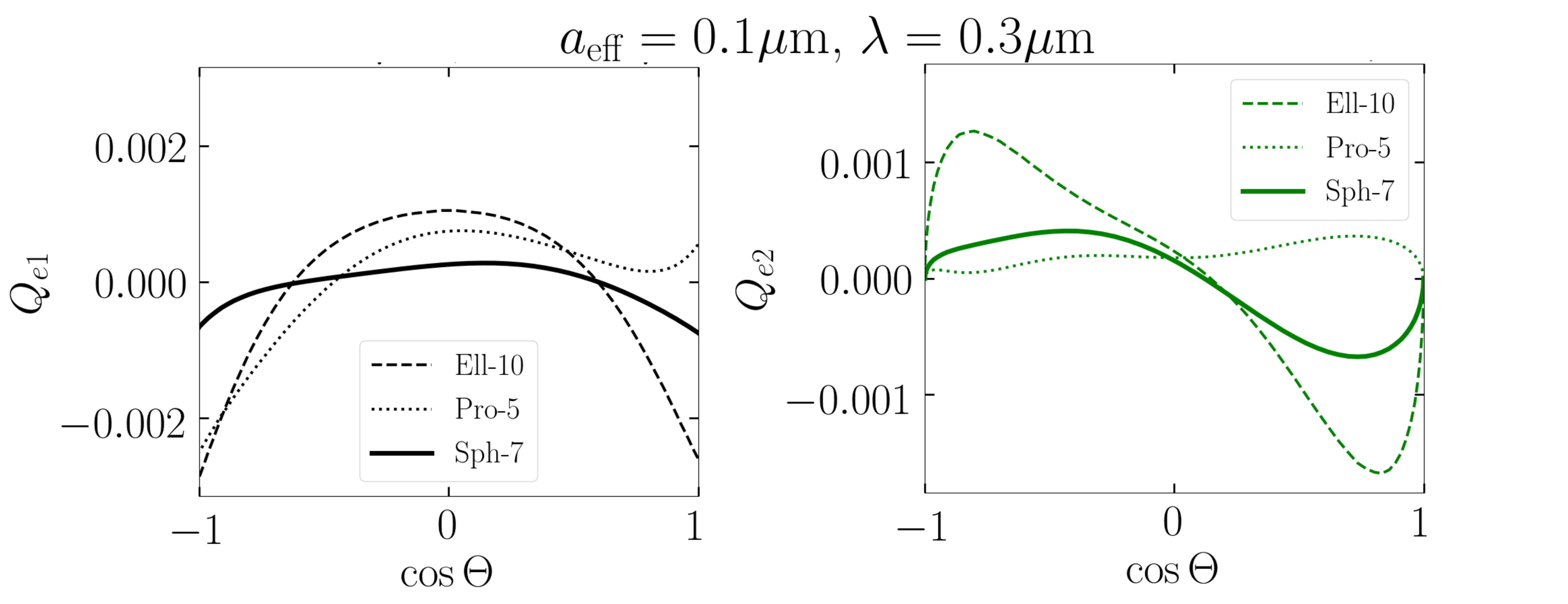}
	\caption{RAT components as in Figure \ref{fig:1200nmRAT-lhand}, but for $ \lambda=0.3 $ $ \mu $m with grains exhibiting mostly left-handed helicity. The prolate grain 5 transitions to right-handed helicity when the major principal axis is nearly parallel to the incident radiation direction. }
	\label{fig:300nmRAT-lhand}
\end{figure}

\begin{figure}
	\centering
	\includegraphics[width=\linewidth]{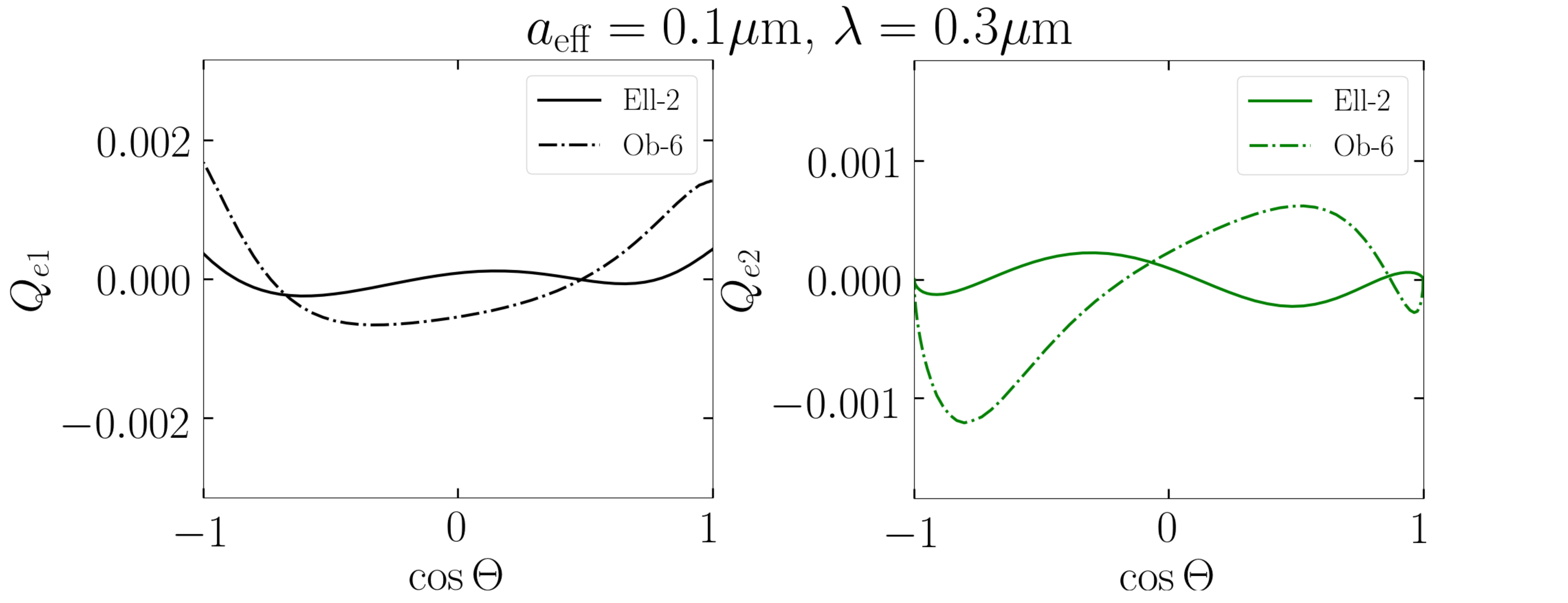}
	\caption{RAT components as in Figure \ref{fig:1200nmRAT-rhand}, but for $ \lambda=0.3 $ $ \mu $m with grains exhibiting a more right-handed helicity. The ellipsoidal grain 2 has helicity transitions when the major principal axis is nearly perpendicular to the incident radiation direction.}
	\label{fig:300nmRAT-rhand}
\end{figure}

To see how the functional form of the RAT components $ Q_{e1} $ and $ Q_{e2} $ behave, we consider five random shapes from the 0.1 $ \mu $m ensemble at $ \lambda = 1.2 $ $\mu$m. The characteristic features, helicity differences and the approximate zeros of the $ \beta $-averaged $ Q_{e2} $ component both appear, as seen in Figures \ref{fig:1200nmRAT-lhand}--\ref{fig:300nmRAT-rhand}. The functional form of RATs deviate from those of the analytical model when size parameter $ x = 2\pi a_{\mathrm{eff}}/\lambda $ grows, as expected. For example, in Figure \ref{fig:1200nmRAT-lhand}, where $ x \approx  0.5$, RATs are practically identical to those of AMO presented in LH07. 

When the size parameter grows, the grain helicities may become more ambiguous as both left- and right-handed helicities are exhibited. This is exemplified in Figures \ref{fig:300nmRAT-lhand} and \ref{fig:300nmRAT-rhand}, which correspond to size parameter $ x \approx 2$. RATs of randomly deformed shapes can deviate clearly from those of AMO in certain cases when $ x>1 $, which was not demonstrated as pronouncingly by the irregular shapes in LH07.

The transition of helicity implies that when grains are both left- and right-handed, different wavelengths may see certain shape features differently. The total handedness is not expected to change for a constant shape. The absolute magnitudes of the RAT components in Figures \ref{fig:1200nmRAT-lhand} and \ref{fig:1200nmRAT-rhand} show that the shapes whose handedness varies with the wavelenght also have the smallest RAT efficiencies. Thus, in these cases even small deviations from AMO can introduce change in the interpreted handedness.

From the visible shapes of the RAT components, we can see that the $ q $-factor varies around unity for many shapes with some outliers such the prolate spheroid in the $ \lambda = $ 0.3 $ \mu $m case. It is cumbersome to analyse the functional forms of a larger amount of particles at once. Thus, in the next section, we probe the statistical behaviour of the $ q $-factor for the whole ensemble.

\section{Analysis of RATs for the ensemble}\label{sec:analysis}
An ensemble of randomly deformed basic shapes allows the collection of q-factor statistics. The shape and composition both affect the distribution of the q-factor. This in turn determines, how effective the alignment without superparamagnetic inclusions will be for different ensembles. In this section, we find distributions of the q-factor for differently shaped and composed ensembles.

\subsection{Effect of shape}
\begin{figure}
	\centering
	\includegraphics[width=\linewidth]{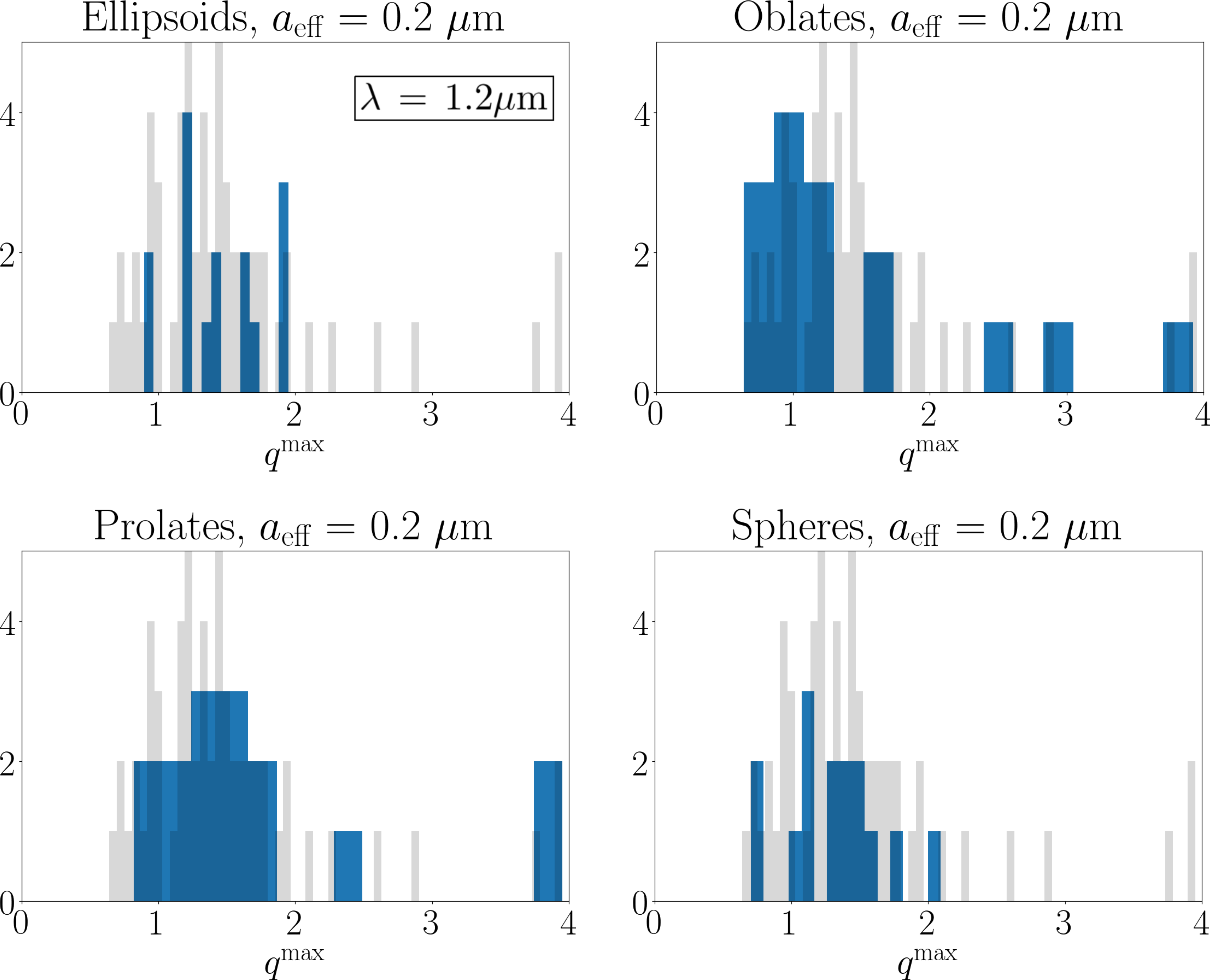}
	\caption{The $ q $-factor distributions of different shapes overlaid with the total distribution of all shapes. The major spread for all particles are approximately the same, with the spheroids deviating the most.}
	\label{fig:1200nmhist}
\end{figure}

First, we compare the effect of shape on the 0.2 $ \mu $m ensemble with $ \lambda = 1.2 $ $\mu$m. The distributions are illustrated in Figure \ref{fig:1200nmhist}. All distributions besides of the prolate shape are centered around $ q =  $ 1.3--1.4, with the prolate distribution centered at $ q = 1.75 $. Statistics of this distribution are colleced in Table \ref*{table:1200nmshape}.

\begin{table}
	\centering
	\caption{The mean, median and standard deviation for the 0.2 $ \mu $m ensemble with $ \lambda = 1.2 $ $ \mu $m for each shape. For each row except total, $ N = 15 $.}
	\begin{tabular}{r|ccc}
		\textbf{Shape}     &\textbf{ Mean}   &\textbf{ Median} & \textbf{STD}    \\
		\hline
		Ellipsoid & 1.4461 & 1.4140 & 0.3333 \\
		Oblate    & 1.4925 & 1.2450 & 0.8913 \\
		Prolate   & 1.7583 & 1.4842 & 0.8926 \\
		Sphere    & 1.308  & 1.3228 & 0.3528 \\
		\hline \hline
		\textbf{Total}	  & 1.5013 & 1.3601 & 0.6952
	\end{tabular}
	\label{table:1200nmshape}
\end{table}

Distributions of $ q^{\mathrm{max}} $ such as in Figure \ref{fig:1200nmhist} would constrain the grain dynamics in the ISRF considerably. As presented in LH07, $ q^{\mathrm{max}} \in (0.5,2) $ indicates that high-J attractors are unlikely. In this case, a high degree of alignment would be resulted only by enhancement effects such as superparamagnetism.

\begin{figure}
	\centering
	\includegraphics[width=\linewidth]{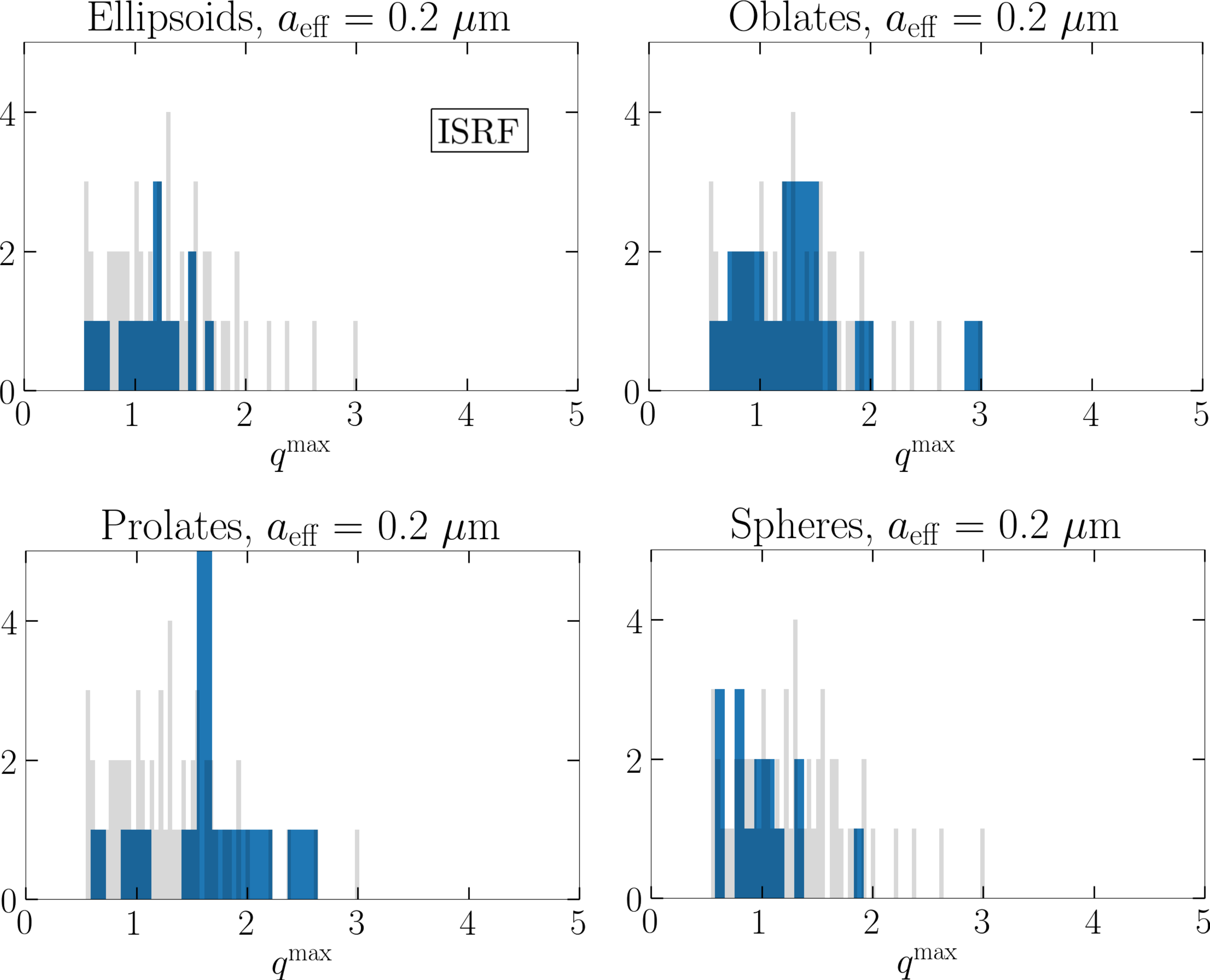}
	\caption{The same as in Figure \ref{fig:1200nmhist}, but for the ISRF spectrum.}
	\label{fig:2m7-isrfhist}
\end{figure}

Motivated by above, we also consider all differently sized ensembles under ISRF illumination. The simplified ISRF spectrum consists of a 5800 K black body intensity spectrum, 10 wavelengths linearly spaced between 120--1740 nm with a peak in the 120 nm ultraviolet component. The ultraviolet peak is 1/4th of the peak of the black body spectrum. The RAT efficiencies are scaled relative to these intensities when determining the $ q^{\mathrm{max}} $ values. For these parameters, the 0.2 $ \mu $m distributions other than that of prolate shapes are centered around unity. Again, the prolate distribution is centered closer to 2, as is illustrated in Figure \ref{fig:2m7-isrfhist}.

\begin{figure}
	\centering
	\includegraphics[width=\linewidth]{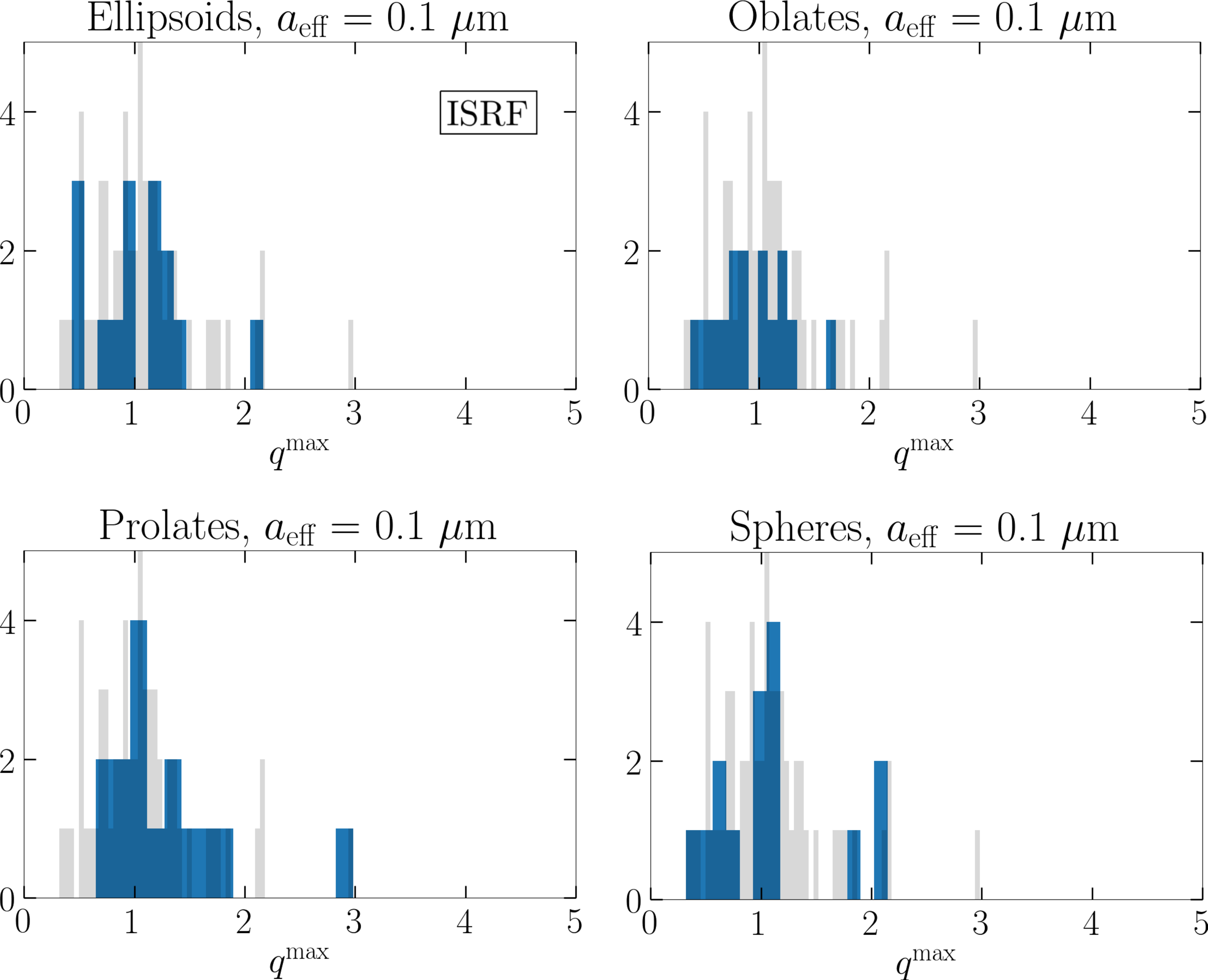}
	\caption{The same as in Figure \ref{fig:2m7-isrfhist}, but for the 0.1 $ \mu $m ensemble.}
	\label{fig:1m7-isrfhist}
\end{figure}

\begin{figure}
	\centering
	\includegraphics[width=\linewidth]{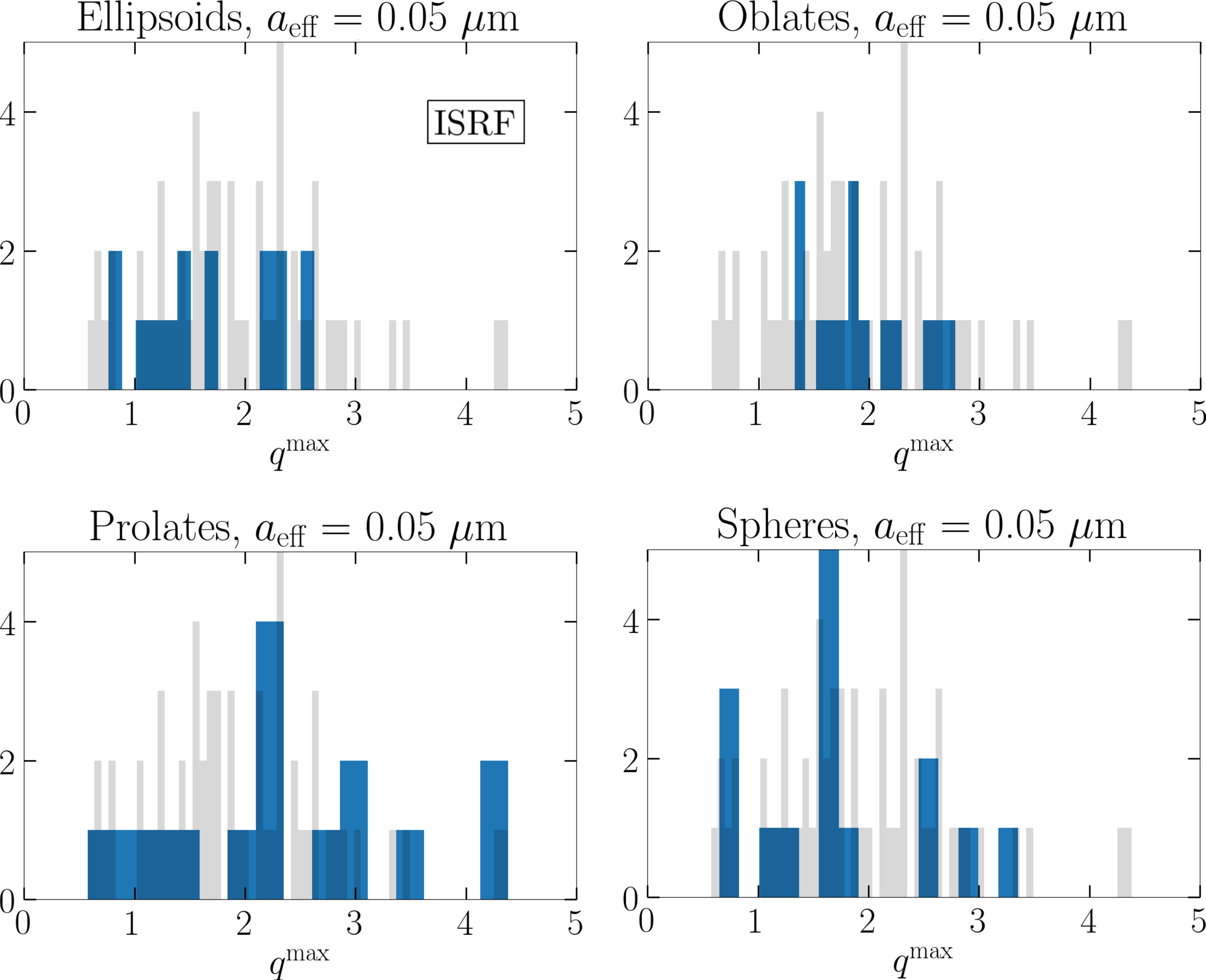}
	\caption{The same as in Figure \ref{fig:2m7-isrfhist}, but for the 0.05 $ \mu $m ensemble.}
	\label{fig:5m8-isrfhist}
\end{figure}

In Figures \ref{fig:1m7-isrfhist} and \ref{fig:5m8-isrfhist}, the 0.1 and 0.05 $ \mu $m ensembles are considered. A similar trend of the prolate shape having distributions centered more right are observed. However, the 0.05 $ \mu $m ensemble has the most spread out distributions and all distributions centered more rightwards than the larger grain ensembles. The smallest grains are expected to exhibit AMO-like RATs in the visible and IR wavelenghts. Thus, only the 120 nm and 300 nm wavelengths are most likely to give rise to irregular RATs, seen e.g. in Figure \ref{fig:300nmRAT-rhand}. This would, after averaging the RATs, result in larger spread of $ q^{\mathrm{max}} $ for smaller grains, as is now seen in Figure \ref{fig:5m8-isrfhist}. The statistics of these ISRF distributions are collected in Table \ref{table:isrfshape}.

\begin{table}
	\centering
	\caption{As in Table \ref{table:1200nmshape}, but for all sizes and with the simplified ISRF spectrum. For each row except total, $ N = 15 $.}
	\begin{tabular}{rr|ccc}
		\textbf{Size}($ \mu $m)     &\textbf{Shape}     &\textbf{ Mean}   &\textbf{ Median} & \textbf{STD}    \\
		\hline
		0.05 & Ellipsoid & 1.7040 & 1.6783 & 0.5983 \\
		0.05 & Oblate    & 1.9307 & 1.8740 & 0.4365 \\
		0.05 & Prolate   & 2.4061 & 2.3276 & 1.0637 \\
		0.05 & Sphere    & 1.7012 & 1.6471 & 0.7706 \\
		\hline \hline
		& \textbf{Total}	  & 1.9355 & 1.8117 & 0.8068 \\
		\hline
		0.1 & Ellipsoid & 1.0283 & 0.9496 & 0.4228 \\
		0.1 & Oblate    & 0.9438 & 0.8650 & 0.3294 \\
		0.1 & Prolate   & 1.2758 & 1.0824 & 0.5571 \\
		0.1 & Sphere    & 1.0935 & 1.0460 & 0.5351 \\
		\hline \hline
		& \textbf{Total}	  & 1.0854 & 1.0511 & 0.4857 \\
		\hline
		0.2 & Ellipsoid & 1.1445 & 1.2027 & 0.3188 \\
		0.2 & Oblate    & 1.3326 & 1.3023 & 0.5737 \\
		0.2 & Prolate   & 1.6617 & 1.6593 & 0.5214 \\
		0.2 & Sphere    & 0.9882 & 0.9487 & 0.3338 \\
		\hline \hline
		& \textbf{Total}	  & 1.2817 & 1.2217 & 0.5162
	\end{tabular}
	\label{table:isrfshape}
\end{table}

\subsection{Effect of composition}
In the total ensemble, there exists five types of grains: pyroxene and olivine type silicates \citep{Scott1996} with thin carbon mantle \citep{Jones2012}, silicates with thicker carbon mantle, variants of previous grains with 7\% iron inclusions \citep{Ordal1983,Ordal1985,Ordal1988} in the silicate core, and amorphous carbon grains. In general, the iron silicates push the real part of the refractive index higher and widen the imaginary part peak in the ultraviolet. Furthermore, amorphous carbon has a naturally high and wide absorption peak in its refractive index. All analysis in this section is done using the aforementioned simplified ISRF spectrum.

\begin{figure}
	\centering
	\includegraphics[width=\linewidth]{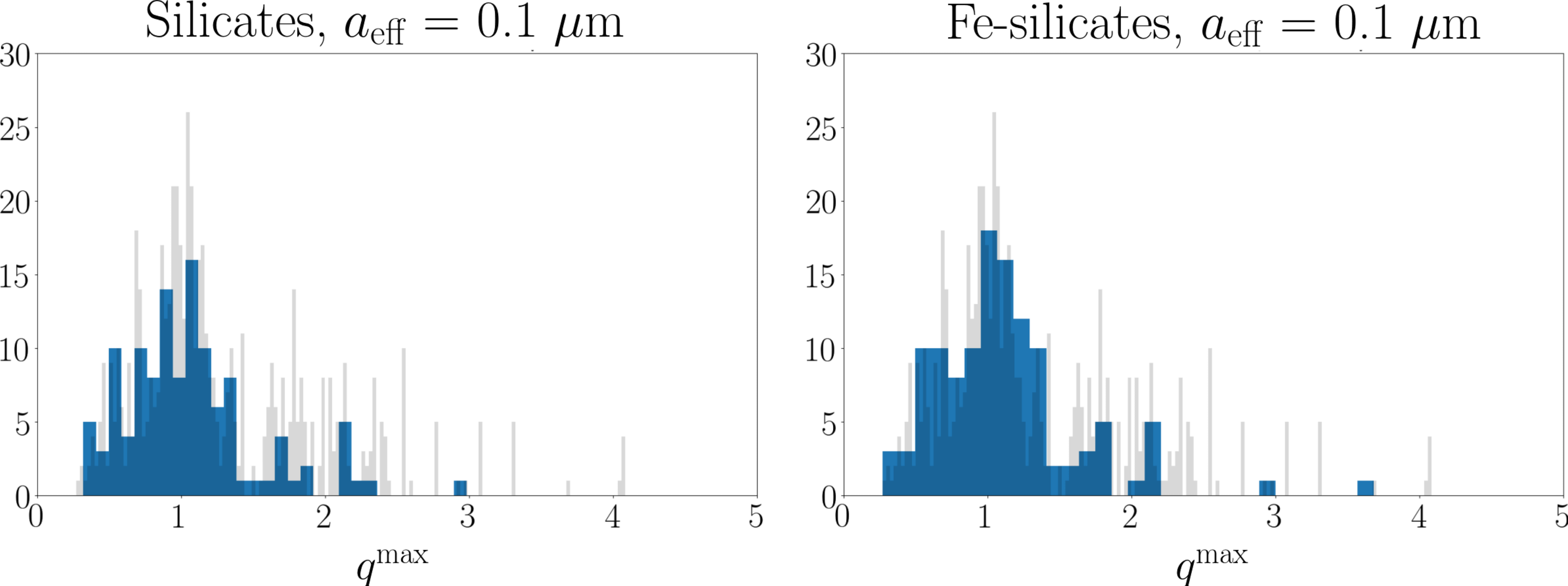}
	\caption{Comparison between $ q^{\mathrm{max}} $ of 0.1 $ \mu $m silicates with and without iron inclusions. The center of distributions are approximately equal, with only slight differences between distributions. }
	\label{fig:ironcomp}
\end{figure}

While iron inclusions severely affect the superparamagnetic properties of grains, they do not significantly change the calculated $ q $-factors. This is demonstrated in Figure \ref{fig:ironcomp}, where the distributions of grains with and without inclusions are nearly identical. Thus, all silicates can be differentiated only by their mantles in this study, leaving three main groups: silicate, carbon-coated silicate, and fully carbonaceous grains.

\begin{figure*}
	\centering
	\includegraphics[width=\linewidth]{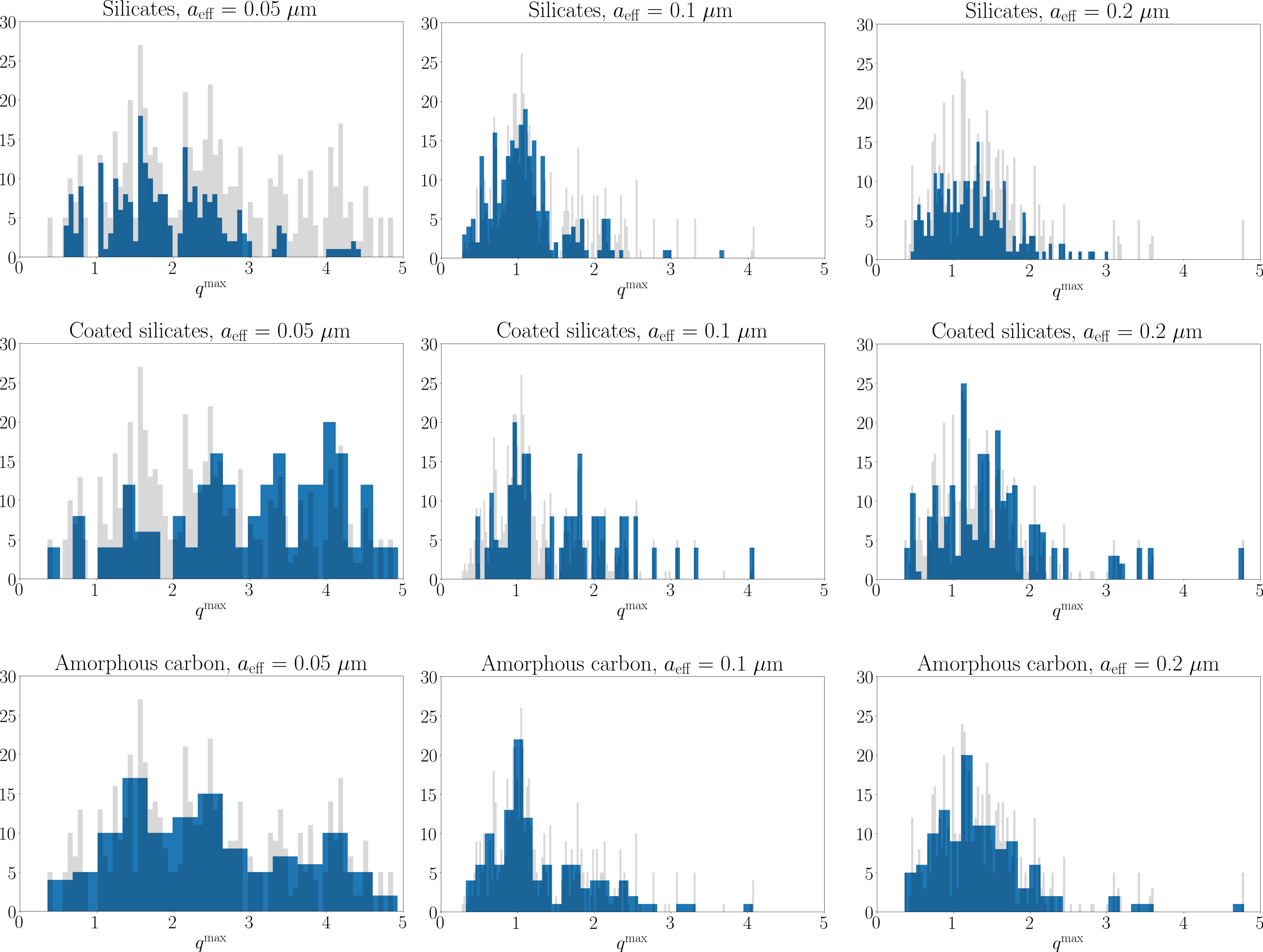}
	\caption{Comparison of $ q^{\mathrm{max}} $ between composition groups of all sizes in the ISRF.}
	\label{fig:allcomp}
\end{figure*}

Comparison between these grain groups in Figure \ref{fig:allcomp} reveals a significant change of $ q^{\mathrm{max}} $ distribution for the 0.05 $ \mu $m ensemble when amount of carbon is large. However, in the case of larger particles, the effect is less notable, while the tendency of the distribution to shift to larger values when adding carbon still exists. This is again probably due to the averaging of more limited amount of RATs with irregular behaviour. In order to constrain the possible compositions of grains in observations, a more complete study over wavelenghts is needed. Indeed, if such spread of $ q^{\mathrm{max}} $ values would persist in a more realistic incident fields, alignment degrees could in some cases constrain the compositions. 

The most probable $ q $-factor falls in the range $ q^{\mathrm{max}} = $ 1.1--1.5 for the two larger grain ensembles regarless of composition and varies from 1.9 to 3.3 for the 0.05 $ \mu $m ensemble. Total statistics are collected in Table \ref{table:isrfcomp}. 

\begin{table*}
	\centering
	\caption{As in Table \ref{table:1200nmshape}, but for the three different composition groups and with the simplified ISRF spectrum.}
	\begin{tabular}{rr|cccc}
		\textbf{Size}($ \mu $m)     &\textbf{Composition}     &\textbf{ Mean}   &\textbf{ Median} & \textbf{STD}   & \textbf{N}  \\
		\hline
		0.05 & Silicate & 1.9214 & 1.7963 & 0.7920  & 240 \\
		0.05 & Coated    & 3.3230 & 3.3197 & 1.6573 & 240 \\
		0.05 & Carbon   & 2.6158 & 2.4175 & 1.4760 & 120 \\
		\hline \hline
		& \textbf{Total}	  & 2.6209 & 2.4186 & 1.4759 & 600 \\
		\hline
		0.1 & Silicate & 1.0886 & 1.0466 & 0.4914 & 240 \\
		0.1 & Coated    & 1.5437 & 1.4319 & 0.7470 & 240 \\
		0.1 & Carbon   & 1.2968 & 1.0718 & 0.6679 & 120 \\
		\hline \hline
		& \textbf{Total}	  & 1.3123 & 1.0942 & 0.6712 & 600 \\
		\hline
		0.2 & Silicate & 1.2488 & 1.1923 & 0.4880 & 240 \\
		0.2 & Coated    & 1.5360 & 1.4386 & 0.8031 & 240 \\
		0.2 & Carbon   & 1.3485 & 1.2179 & 0.6651 & 120 \\
		\hline \hline
		& \textbf{Total}	  & 1.3836 & 1.2920 & 0.6771 & 600
	\end{tabular}
	\label{table:isrfcomp}
\end{table*}

\subsection{Comparison with AMO}
Using the ensemble, a more extensive RAT component comparison between arbitrary grain shapes and AMO can be obtained for justification of the model. We thus proceed similarly as in LH07.

\begin{figure}
	\centering
	\includegraphics[width=\linewidth]{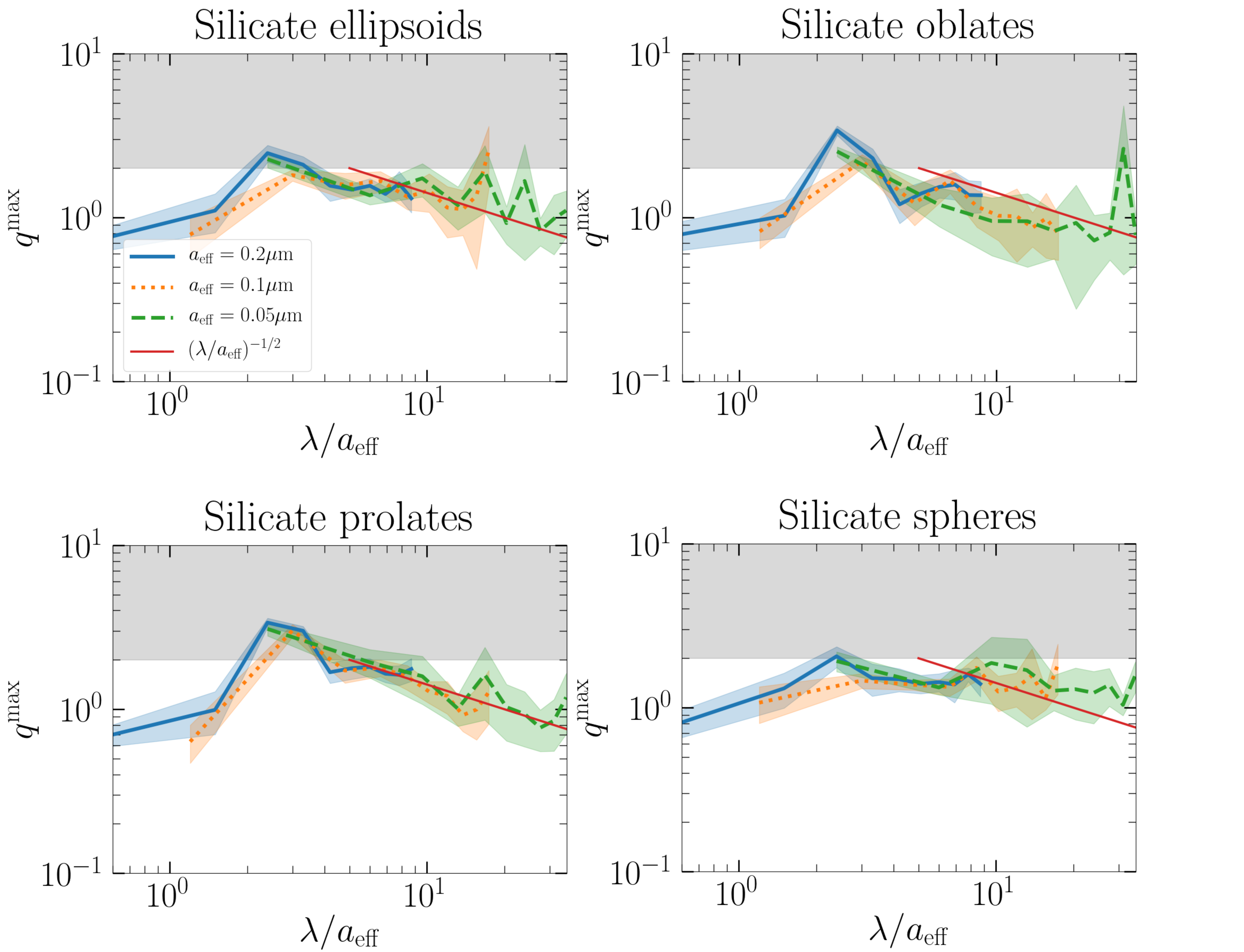}
	\caption{$ q^{\mathrm{max}} $ as the function of $ \lambda/a_{\mathrm{eff}} $ for the silicate base shapes. The coefficient of variation is indicated by the shaded area around the mean. This statistic adequately corresponds to the amount of individual $ q^{\mathrm{max}} $ values lying in and out of the darker area where high-$ J $ alignment is possible.}
	\label{fig:qmax-sil}
\end{figure}

\begin{figure}
	\centering
	\includegraphics[width=\linewidth]{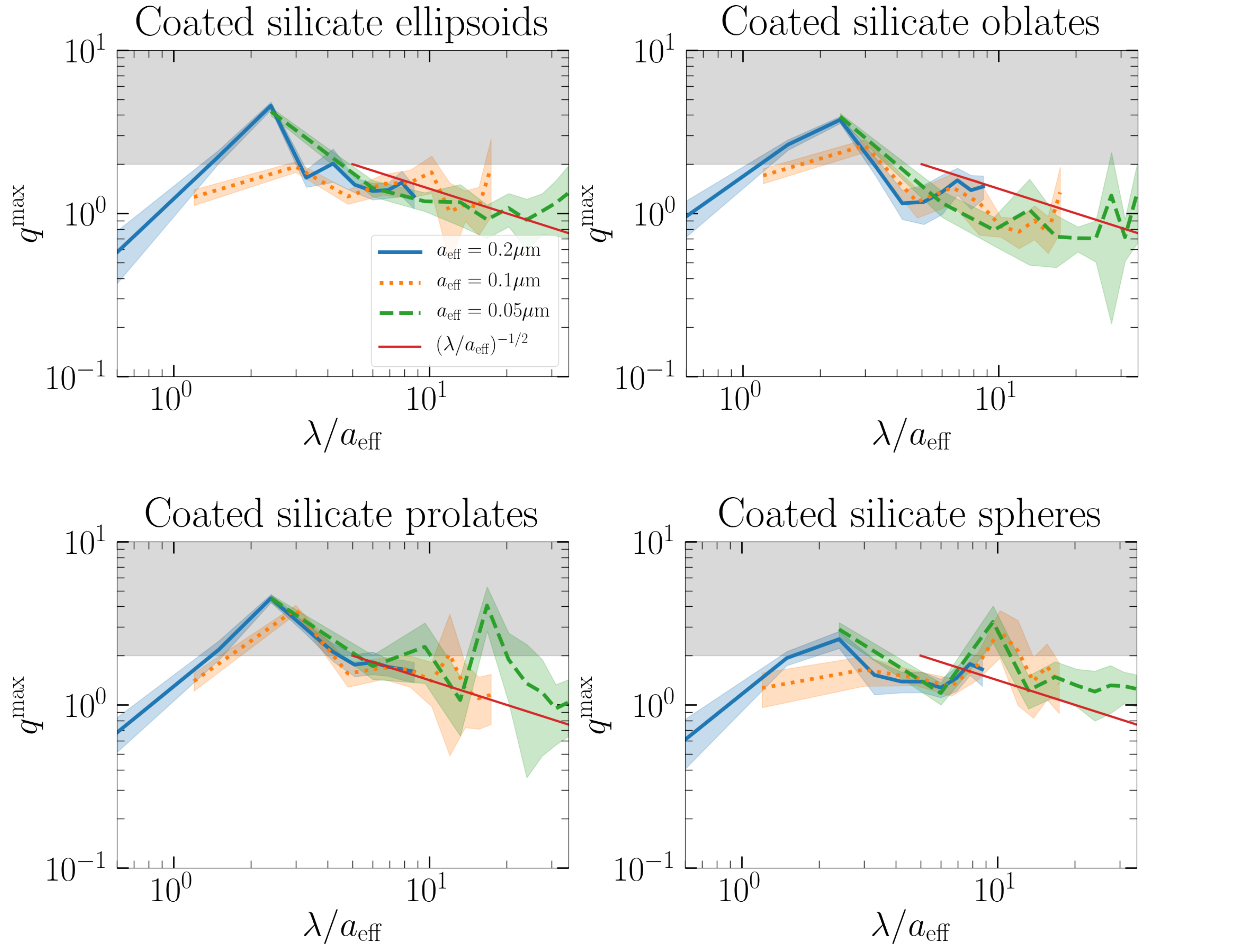}
	\caption{The same as in Figure \ref{fig:qmax-sil}, but for coated silicates.}
	\label{fig:qmax-csil}
\end{figure}

\begin{figure}
	\centering
	\includegraphics[width=\linewidth]{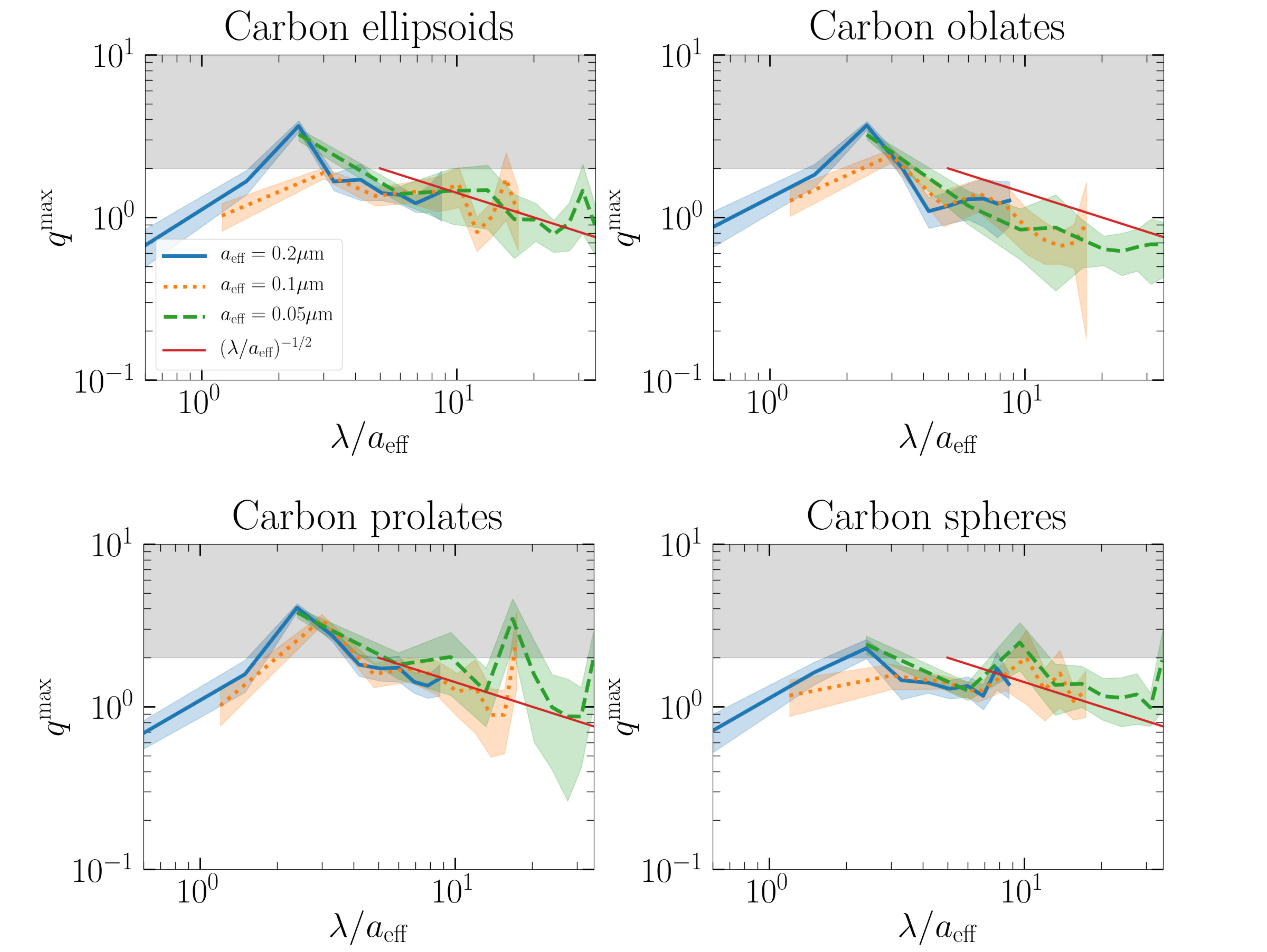}
	\caption{The same as in Figure \ref{fig:qmax-sil}, but for carbonaceous shapes.}
	\label{fig:qmax-C}
\end{figure}

First, we compare the effect of both shape and composition on $ q_{\mathrm{max}} $ as a function of $ \lambda/a_{\mathrm{eff}} $, or the ratio of wavelength and effective size of the grain. The mean values and the coefficient of variation  of $ q^{\mathrm{max}} $ are illustrated in Figures \ref{fig:qmax-sil} -- \ref{fig:qmax-C}. 

In almost all the cases, the 0.1 $ \mu $m grains have peaks in $ q^{\mathrm{max}} $ near $ \lambda/a_{\mathrm{eff}} = 3$. Namely, only the peak of the prolate shape, regardless of shape, reaches the high-$ J $ values. The general downward trend is proportional to $ (\lambda/a_{\mathrm{eff}})^{-1/2} $. Both carbon-coated silicate grains and purely carbonaceous grains also exhibit some relatively strong deviations from the trend for both prolate spheroids and spherical shapes. These data complement the observations in e.g. Figure \ref{fig:allcomp}, where pure 0.05 $ \mu $m silicates have a highly different distribution from ensembles containing carbon, as can be seen from the difference peaking of 0.05 $ \mu $m data between Figures \ref{fig:qmax-sil} -- \ref{fig:qmax-C}.

Second, we perform $ \langle \Delta^{2} \rangle $ testing between the ensemble and AMO, where $ \langle \Delta^{2} \rangle $ gives the mean deviation of RAT components over $ \Theta $ as in LH07: 
\begin{equation}
\langle \Delta^{2} \rangle (Q_{ei}) = \dfrac{1}{\pi(Q_{ei}^{\mathrm{max}})^{2}} \int_{0}^{\pi}[Q_{ei}^{\mathrm{irreg}}-Q_{ei}^{\mathrm{AMO}}]^{2}\mathrm{d}\Theta.
\end{equation}

\begin{figure}
	\centering
	\includegraphics[width=\linewidth]{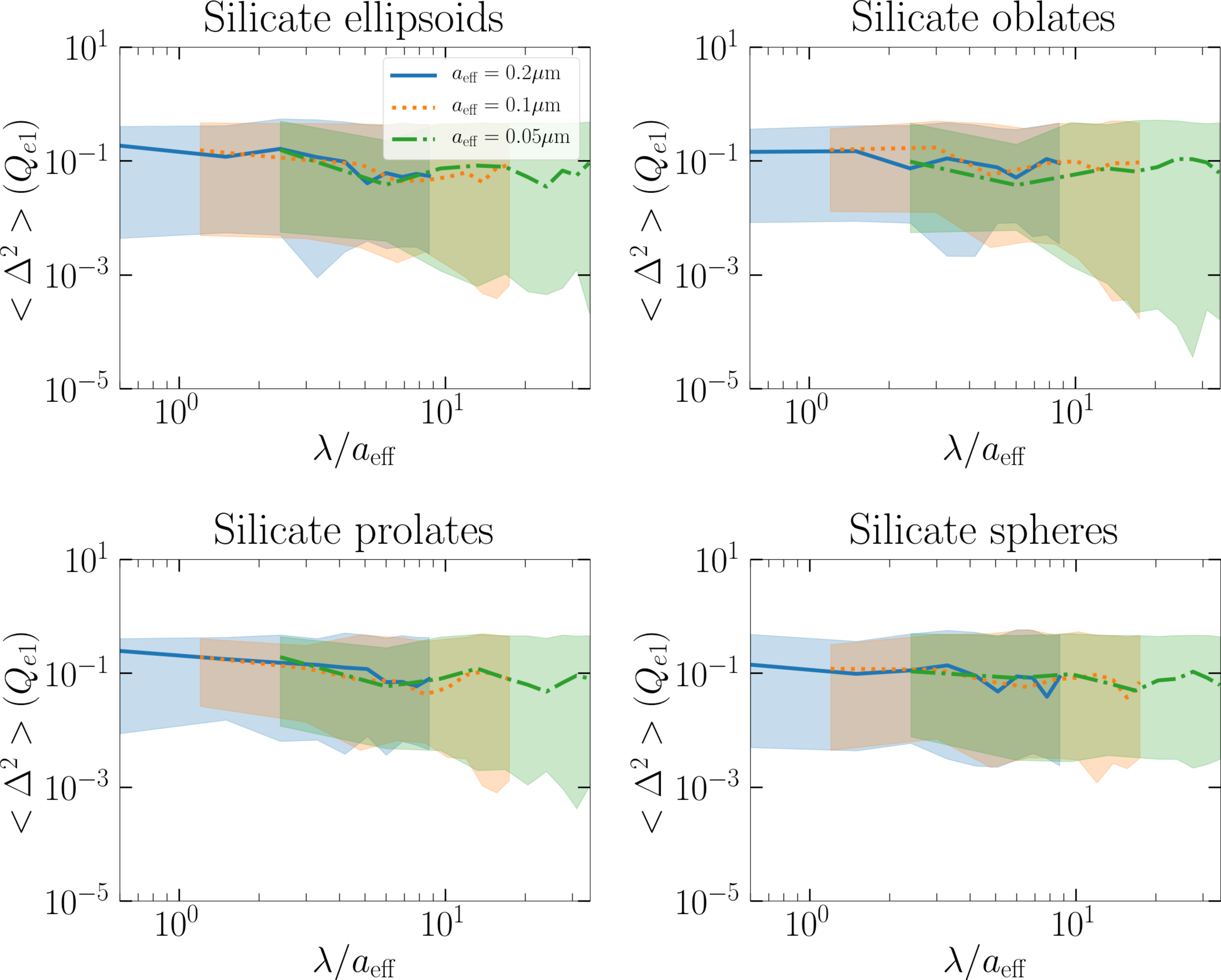}
	\caption{Mean deviation and range of the $ Q_{e1} $ component between each irregular silicate shape ensemble and AMO.}
	\label{fig:qdiff-e1}
\end{figure}

\begin{figure}
	\centering
	\includegraphics[width=\linewidth]{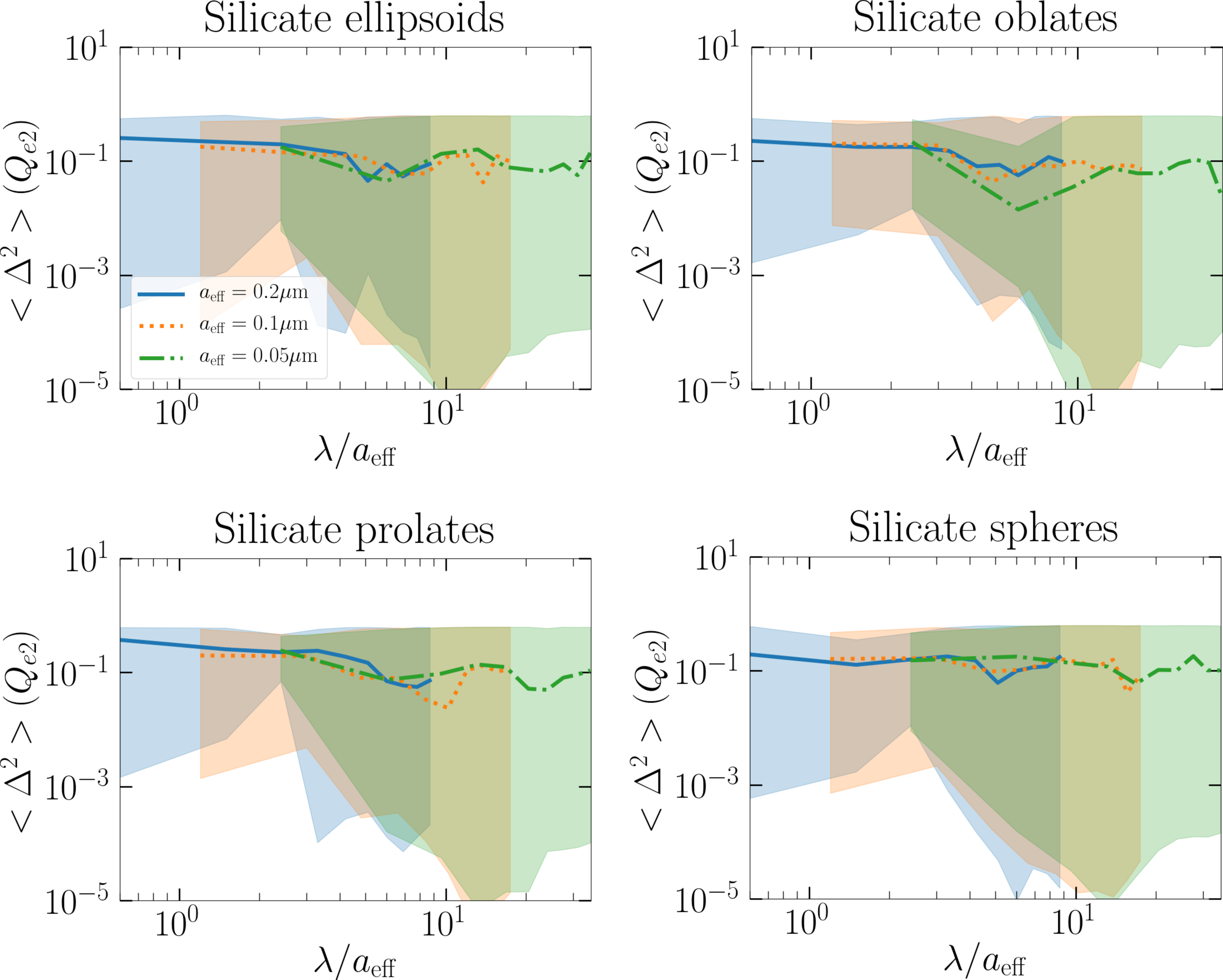}
	\caption{The same as in Figure \ref{fig:qdiff-e1}, but for the $ Q_{e2} $ component.}
	\label{fig:qdiff-e2}
\end{figure}

Grain shape has a large effect on the deviation for both RAT components, as are illustrated in Figures \ref{fig:qdiff-e1} and \ref{fig:qdiff-e2} for silicates only, as composition does not visibly affect the deviation distribution. The mean value can be seen to decrease from 10--20\% to few percent as grains get smaller. The total range of the deviations vary largely, $ Q_{e1} $ ranges from $ 10^{-3} $ ($ 10^{-4} $ for oblate spheroids) to $ 4\times10^{-1} $. Similar is true for $ Q_{e2} $ with even less variations between the different base shapes.

The range of difference from AMO in Figure \ref{fig:qdiff-e2} are considerably larger than in LH07. This was already hinted in Figures \ref{fig:300nmRAT-lhand} and \ref{fig:300nmRAT-rhand}, where the functional form of $ Q_{e2} $ are visibly more irregular than those presented in LH07. However, such behaviour is natural when larger set of randomly deformed irregular shapes are considered. 

Third, in the AMO framework, RATs are not expected to depend strongly on the grain composition, but on grain helicity. Our calculations indeed show the similar magnitude of RATs for the different compositions.

\begin{figure}
	\centering
	\includegraphics[width=\linewidth]{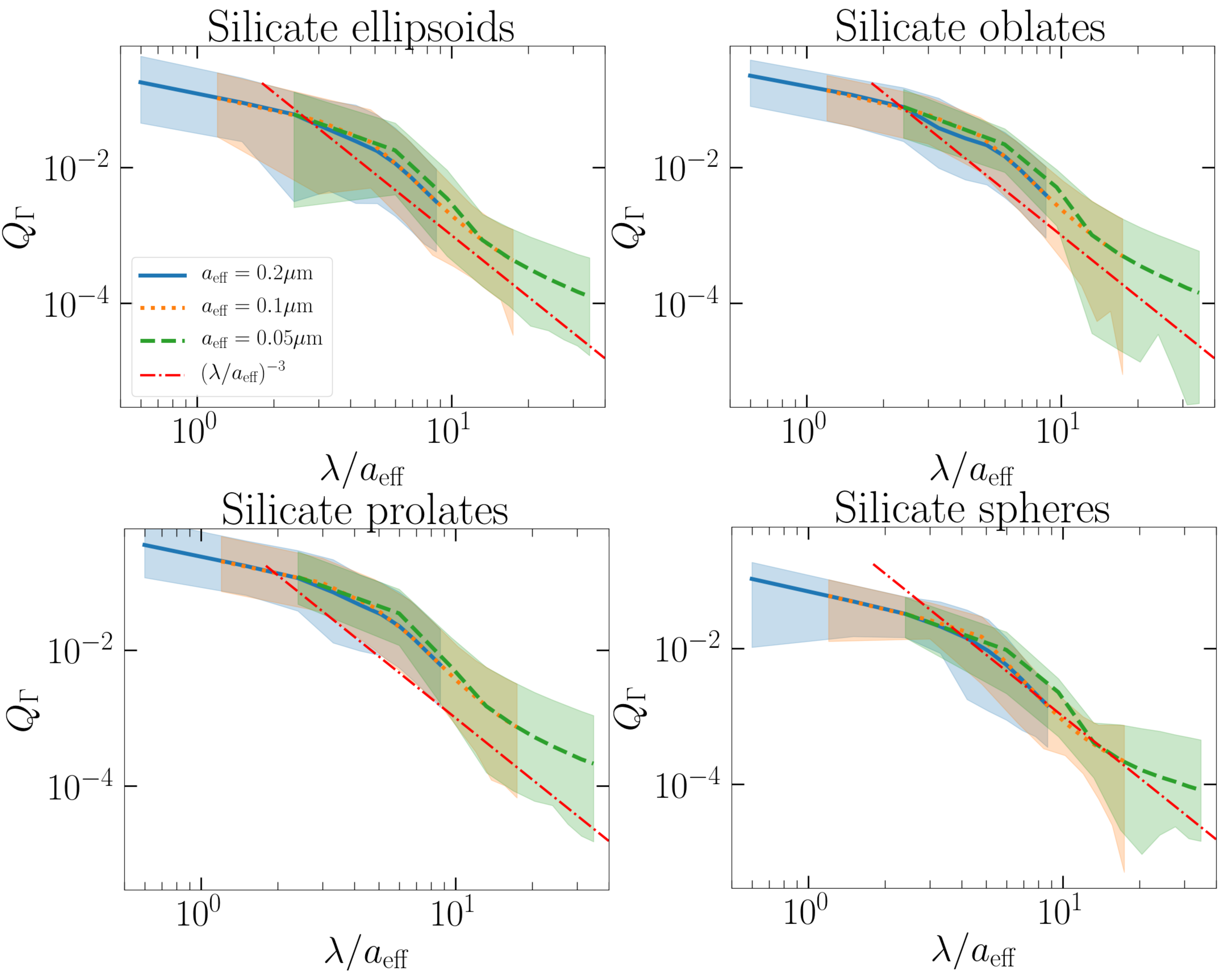}
	\caption{$ Q_{\Gamma}(\Theta=0) $ as the function of $ \lambda/a_{\mathrm{eff}} $ for the silicate base shapes. The shaded area around the mean indicates the range of all samples.}
	\label{fig:Qt-sil}
\end{figure}

\begin{figure}
	\centering
	\includegraphics[width=\linewidth]{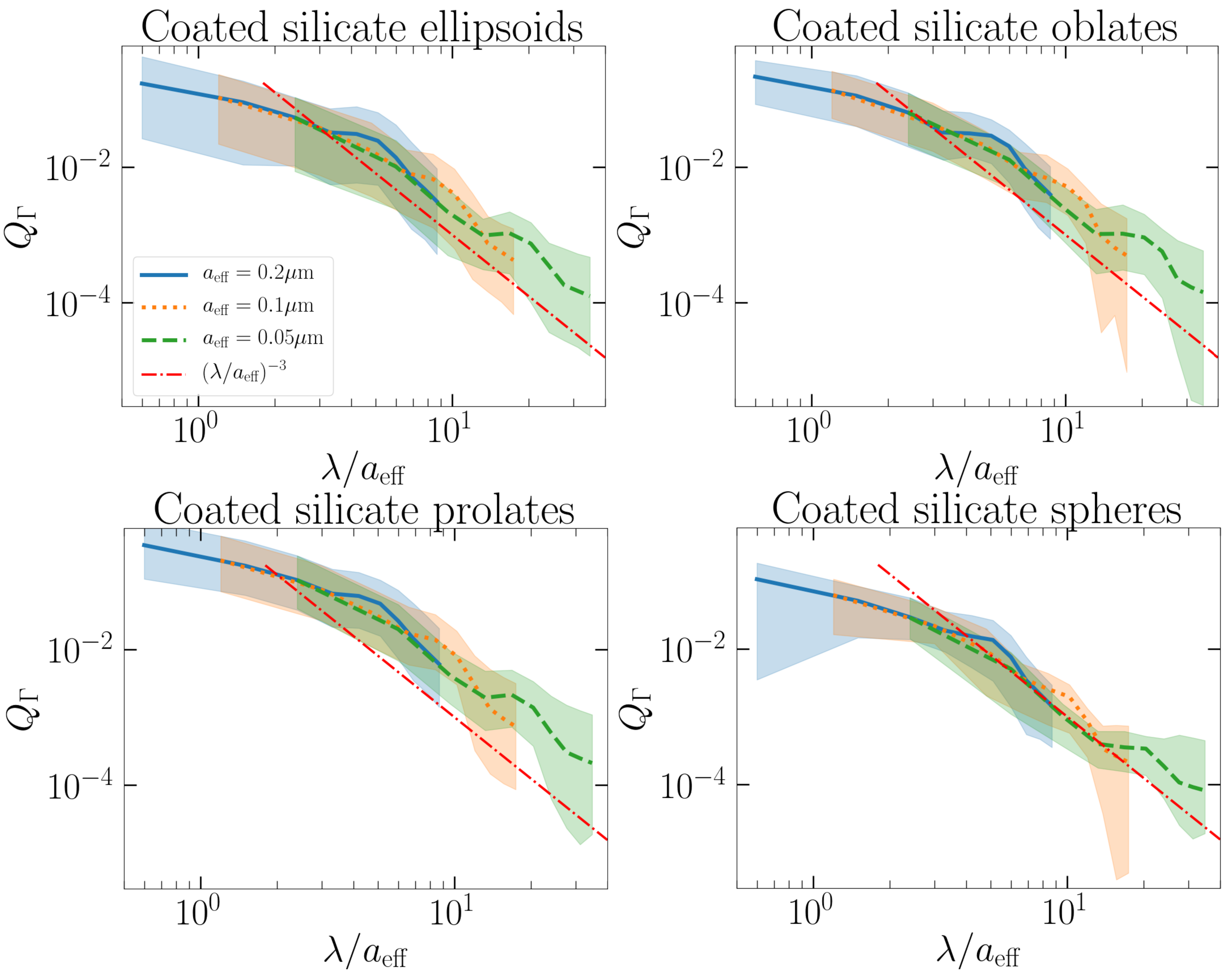}
	\caption{The same as in Figure \ref{fig:Qt-sil}, but for coated silicates.}
	\label{fig:Qt-csil}
\end{figure}

\begin{figure}
	\centering
	\includegraphics[width=\linewidth]{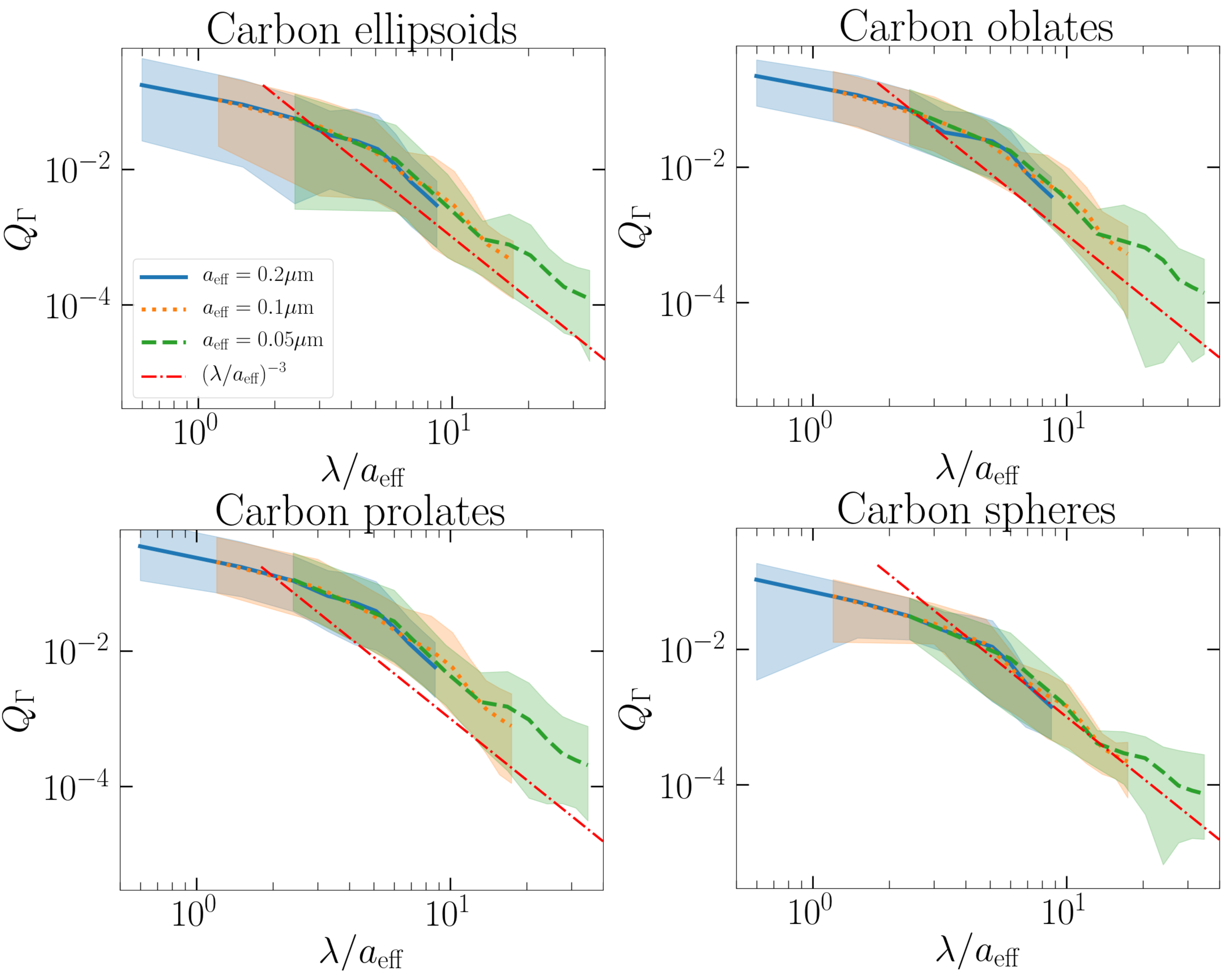}
	\caption{The same as in Figure \ref{fig:Qt-sil}, but for carbonaceous shapes.}
	\label{fig:Qt-C}
\end{figure}

Lastly, we consider the scaling predictions of torques given by the ensemble. The RAT magnitudes $ Q_{\Gamma} $ at $ \Theta = 0 $ were calculated. Mean values and ranges of $ Q_{\Gamma} $ are presented in Figures \ref{fig:Qt-sil} -- \ref{fig:Qt-C}. In all cases, self-similarity as reported in LH07 (e.g. Figure 30 in LH07) can be observed, moreso in grains containing carbon. After a gradual steepening, all the curves are closely proportional to $ (\lambda/a_{\mathrm{eff}})^{-3} $. This overall best fit agrees well with the analysis in LH07, where the same spectral index $ -3 $ provides the best fit for a severely more limited set of shapes.

Due to the limited wavelenght range available, Figures \ref{fig:Qt-sil} -- \ref{fig:Qt-C} cannot show whether or not the RAT efficiency is best fitted by a constant below $ \lambda/a_{\mathrm{eff}} = 1.8 $. The region $ \lambda \sim a_{\mathrm{eff}} $ is not possible to account for properly using the limited wavelenght range. Due to this, the full RAT fit provided in LH07 is not possible to confirm for the ensembles. 

\section{Discussion}\label{sec:discussion}

\subsection{Importance of the present study}

The absence of constrains on $q^{\mathrm{max}}$ in the theory of RAT alignment limits the predictive power of the theory. For instance, Figure \ref{fig:deltam_cri} (see gray shaded regions) indicates that for ordinary paramagnetic grains the high-J attractor point and therefore the perfect alignment of grains with magnetic field is possible if the $q^{\mathrm{max}}$ parameter is above $2$ for the angle between the radiation and the magnetic field $\psi<45$ degrees and $q^{max}<1$ for $\psi>45$ degrees.

Our present study shows that for most grain shape distribution that we explored the most probable range of $q^{\mathrm{max}}$ is within 1.1-1.5 for large grains (see Section 4.2). This means that the observational detection of perfect alignment, e.g., by {\it Planck} \footnote{In practice it may be easier to search for the variations of the alignment degree with the angle between the magnetic field and the radiation anisotropy direction as it is done in \citet{Andersson2011}. Within the RAT theory the absence of variations of the alignment as revealed by \citet{Planck2018} would indicate the perfect alignment of grains for all the angles and therefore the enhanced magnetic dissipation within grains.} in this range necessarily implies that the alignment arises from the joint action of RATs and magnetic relaxation torques \citep[see][]{Hoang2016} (see orange shaded region in Figure \ref{fig:deltam_cri}). Moreover, it can also constrain the magnetic response of the grain material \citep[see][]{Lazarian2018}. 

\subsection{Role of $q^{\mathrm{max}}$ for grains with enhanced magnetic susceptibilities}

The original model in LH07 has been significantly extended and elaborated in the subsequent studies. Most important was the exploration of the joint action of the RATs and enhanced magnetic relaxation first described in \citet[henceforth LH08]{Lazarian2008c} and numerically demonstrated in \citet[henceforth HL16]{Hoang2016}. While the torques arising from paramagnetic relaxation within ordinary paramagnetic grains are completely negligible, the torques arising from the dissipation within
a grain with the enhanced magnetic response, e.g., superparamagnetic grain \citep[see][]{Morrish1980}, are shown by LH08 to be important in stabilizing the high-J attractor point. Depending on the value of the parameter 
$\delta_{\mathrm{mag}}=t_{\mathrm{damp}}/t_{\mathrm{mag}}$, where $t_{\mathrm{damp}}$ includes various grain randomization/damping processes \citep[see][and ref. therein]{Lazarian2018}, while $t_{\mathrm{mag}}$ is the time of the magnetic relaxation of a grain rotating perpendicular to the magnetic field direction, the parameter space for the $q^{\mathrm{max}}$ and the cosine of the angle between the magnetic field direction and the radiation anisotropy direction, is changing as shown in Figure 2 from HL16. In the latter study, the range of $q^{\mathrm{max}}$ was unconstrained. However, in view of our present study we are able to show the ranges of parameter corresponding to the given classes of grain shapes. 

Recalling that the alignment with the low angular momentum corresponds to the alignment measure in the range of $20$ or $30$\% \citep[henceforth HL08]{Hoang2008}, i.e., is significantly reduced compared to the perfect alignment of grains in the high-J attractor point,\footnote{The alignment is not close to zero \citep[cf][]{Weingartner2003} due to the gaseous bombardment that induces the diffusion of the phase trajectories of grains in the vicinity of the high-J repellor point (HL08).} one can see that the presence of higher grain magnetic response can significantly increase the alignment measure. 

\begin{figure}
	\centering
	\includegraphics[width=\linewidth]{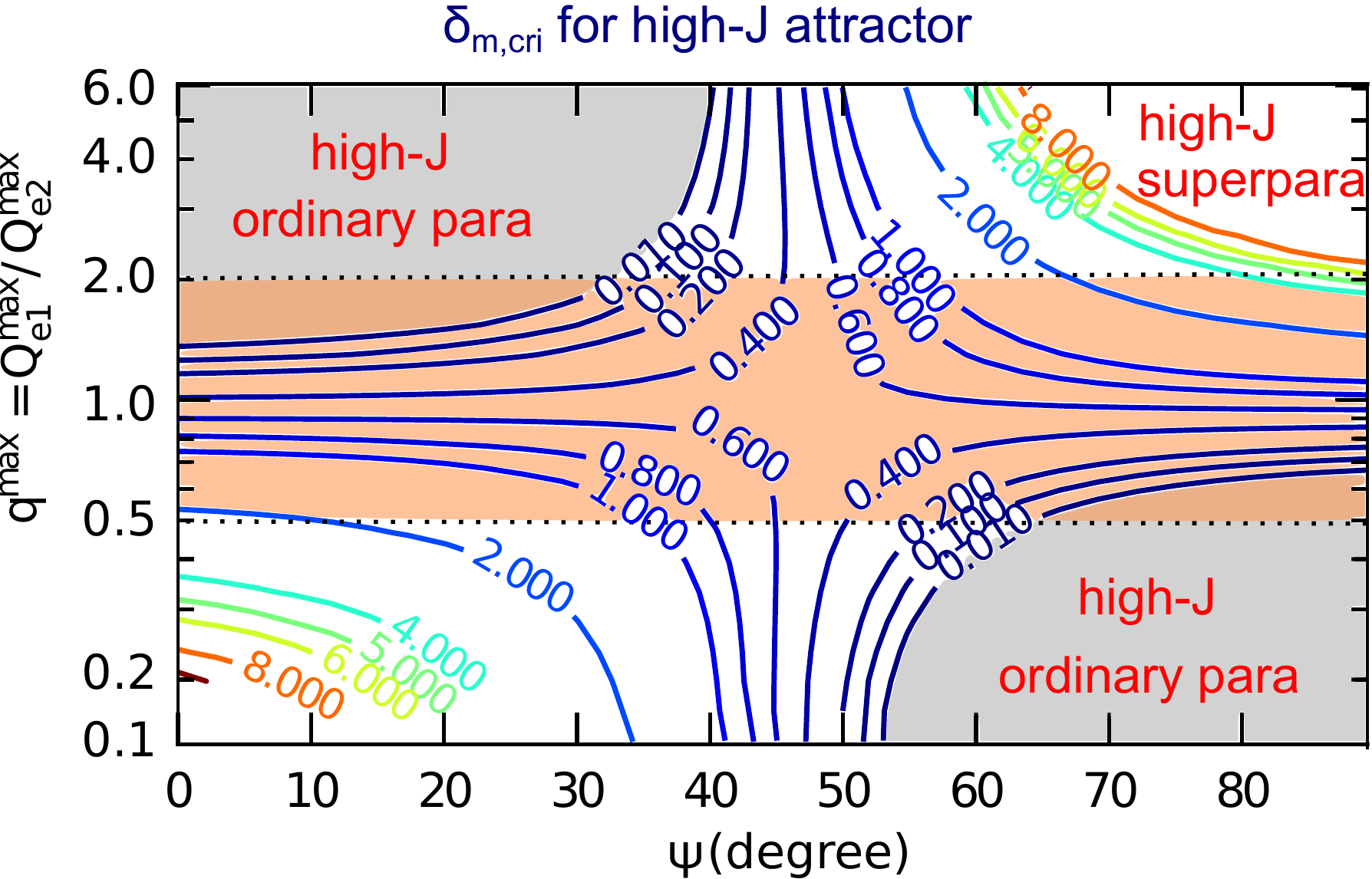}
	\caption{Critical magnetic relaxation parameter $\delta_{m,cri}$ required to have alignment with high-J attractor as functions of $\psi$ and $q^{\mathrm{max}}$. The orange region marks the most probable range of $q^{\mathrm{max}}\sim 0.5-2$ computed for the ensemble of shapes.}
	\label{fig:deltam_cri}
\end{figure}

HL16 performed a detailed numerical studies for grains with varying level of iron inclusions. HL16 derived the critical value of $\delta_{m}$ for which grains can be aligned with high-J attractors and demonstrated that grains with high-J attractors can be perfectly aligned. Figure \ref{fig:deltam_cri} shows the critical relaxation parameter that results in alignment with high-J attractor where the middle region in orange shade corresponds to the most probable range of $q^{\mathrm{max}}$ computed from our grain ensemble. It shows that one only requires $\delta_{m,\mathrm{cri}}<5$ to have alignment with high-J attractors. This reveals that the magnetic susceptibility of grains is enhanced slightly to achieve perfect alignment.

\subsection{Implication for RAT alignment of silicate vs carbonaceous grains}
Our results show that RATs slightly vary with the composition of dust grains. RATs from carbonaceous grains are not much different from silicate grains, which follows that carboncaceous grains can also be spun-up to suprathermal rotation by RATs as silicate and iron grains. As a result, the difference in grain aligment of different grain compositions originates from other physical properties, such as the grain magnetic properties. 
In the diffuse ISM, silicate grains can be aligned with the magnetic field due to fast Larmor precession, but carbonaceous grains are not expected to be aligned with the magnetic field due to slow Larmor precession than the gas collisional randomization, assuming silicate and carbon grains are segregated perhaps due to rotational disruption by RATs \cite{Hoang2018b}. In dense molecular clouds, these two dust populations are mixed together due to grain collisions, thus, one expect the composite siliate and carbon grains can be aligned with the magnetic field.

\subsection{Other effects and the uncertainties in the alignment}

In HL08 the role of suprathermal torques on the RAT alignment was explored. Such torques, e.g. arising from the H$_2$ formation on grain surfaces can raise the low-J attractor point, significantly increasing the composite (RATs + other uncompensated torques) alignment. This effect does not depend on the $q^{\mathrm{max}}$ parameter. 

Other effects explored in earlier studies may also be important, but their role has not been properly quantified yet. For instance, the study that we provide is applicable to classical silicate grains of the size less than $2\cdot10^{-5}$ cm for which the internal relaxation \citep{Purcell1979,Lazarian1999b,Lazarian1999a,Lazarian2018} efficiently aligns the axis of grain rotation and the axis of the maximal moment of grain inertia. For larger grains and also for carbonaceous grains this type of relaxation may not be sufficiently efficient. Then the grains wobble and may get aligned with the grain long axes both parallel and perpendicular to magnetic field. The study in \citet{Hoang2009} identified that internal relaxation of grains larger than 0.6--1.5 $\mu$m becomes inefficient, depending on the intensity of the radiation environment.

\subsection{Other types of alignment}

The alignment of grains can happen due to other processes that are different from the RATs. For instance, if the grain cross section is different in respect to the flow of particles or radiation, the "cross sectional" alignment take place \citep{Lazarian2005}. This mechanism was employed recently in \citet{Hoang2018} to explain the alignment of tiny aromatic carbonaceous grains (PAHs) in the vicinity of the radiation sources. 

The concept of helical grain alignment was generalized to the corpuscular interaction with gaseous atoms \citep{Lazarian2007a}. The relative motion of grains and gas is expected due to turbulence \citep{Lazarian2002a,Yan2002,Yan2004,Hoang2011,Xu2018}. For the toy model of grain in Figure 1, it does not matter whether the interactions are arising from the gas-grain or radiation-grain interactions. However, for realistic irregular grains the interactions are more complex. Unlike the radiation flow that samples the grain entirely and therefore determines the overall grain helicity, the grain helicity that is seen by the gaseous flow depends on the orientation of the grain rotation axes in respect to the flux of the impinging atoms. As a result, numerical simulations in \citet{Hoang2018a} demonstrate the shape of the torques that vary substantially and are not as universal as in the case of the AMO in LH07. Due to the variations of grain helicity as grains shows different facets to the flow, the amplitude of the mechanical torques on helical grains is also reduced. Nevertheless, such torques are more efficient compared to the stochastic mechanical torques associated with the Gold alignment \citep{Gold1952a,Gold1952b,Dolginov1976,Roberge1993,Lazarian1994a,Lazarian1995b}. A more extensive study of the mechanical alignment of helical grains is necessary, but our present approach of calculating the distribution of $q^{\mathrm{max}}$ for different grain shapes does not look promising in this case.

\section{Conclusions}\label{sec:conclusions}

The present study analyzes the distribution of $q^{\mathrm{max}}$ parameter for a few classes of grain shapes that we believe can be present in astrophysical environments. In terms of the number of explored shapes it presents a radical change, i.e. from 5 in LH07 and subsequent studies to 60 in the present study. We confirm that the RATs within the whole variety of shapes explored are consistent with the AMO model predictions in LH07. This, combined with the constrained values of $q^{\mathrm{max}}$ increase our confidence in the RAT theory and increase its predictive power. We found that superparamagnetic inclusions are important in order to have the perfect RAT alignment for a wide variety of shaper that we considered. Our study is important both for studies magnetic fields in interestellar medium as for probing physical conditions in other environments, e.g. comet atmospheres and circumstellar regions. In addition to testing the AMO model, we confirmed empirical relations for the scaling of of RATs with the ratio of the grain size to the wavelength. These relations makes it easier to evaluate the importance of the RAT alignment.

Acknowledgements: AL acknowledges the support by the NSF AST 1715754 grant. TH acknowledges the support from the Basic Science Research Program through the National Research Foundation of Korea (NRF), funded by the Ministry of Education (2017R1D1A1B03035359).

\end{document}